\documentclass[12pt]{article}
\usepackage{color}
\usepackage{amssymb,amsmath,comment}
\usepackage{graphicx}
\usepackage{braket}
\usepackage{epsfig}
\usepackage{here}
\usepackage{bm}
\graphicspath{{./Figures/}}
\newcommand{\ba}{\begin{align}}
\newcommand{\ea}{\end{align}}

\def\ov{\overline}
\def\nn{\nonumber}

\def\bea{\begin{eqnarray}}
\def\eea{\end{eqnarray}}
 \makeatletter
\def\alt{\mathrel{\mathpalette\gl@align<}}
\def\agt{\mathrel{\mathpalette\gl@align>}}
\def\gl@align#1#2{\lower.6ex\vbox{\baselineskip\z@skip\lineskip\z@
\ialign{$\m@th#1\hfil##\hfil$\crcr#2\crcr\sim\crcr}}} \makeatother
\renewcommand{\thefootnote}{\fnsymbol{footnote}}
\begin{document}
\begin{flushright}
\end{flushright}

\rightline{NITEP 279}

\vspace*{1.0cm}

\begin{center}
\baselineskip 20pt 
{\Large\bf 
Optimizing Yukawa couplings to suppress Dimension-five Proton Decay
\\
in $SU(5)$ GUT
}
\vspace{1cm}

{\large 
Naoyuki Haba${}^{a,b}$, Junpei Ikemoto${}^{a}$,  Yasuhiro Shimizu${}^{a,b}$ 
\\
and Toshifumi Yamada${}^{c}$
} \vspace{.5cm}

{\baselineskip 20pt \it
${}^{a}$Department of Physics, Osaka Metropolitan University, Osaka 558-8585, Japan \\
${}^{b}$Nambu Yoichiro Institute of Theoretical and Experimental Physics (NITEP),
Osaka Metropolitan University, Osaka 558-8585, Japan\\
${}^{c}$Institute for Mathematical Informatics, Meiji Gakuin University, Yokohama 244-8539, Japan
}

\vspace{.5cm}

\vspace{1.5cm} {\bf Abstract} \end{center}

The minimal supersymmetric $SU(5)$ grand unified theory (GUT) provides a highly compelling framework for physics beyond the Standard Model (SM). 
However, it suffers from a severe phenomenological challenge: 
 rapid proton decay mediated by colored-Higgsino exchange via dimension-five operators.
Resolving this issue often requires adjustments to the Yukawa couplings and the potential sectors, 
 generating a vast and complex parameter space where traditional brute-force numerical scans are rendered computationally intractable due to the curse of dimensionality. 
In this paper, 
 we overcome this limitation by applying machine learning optimization techniques. 
We investigate a supersymmetric $SU(5)$ model extended with $\mathbf{45}$ and $\overline{\mathbf{45}}$ Higgs representations, 
 defining a loss function based on the partial decay width of $p \to K^+ \bar{\nu}$. 
Utilizing the Adam optimizer,
 we systematically explore the 33-dimensional parameter space to identify regions that suppress proton decay.
Furthermore, we vary $\tan \beta$ to thoroughly investigate whether the optimized proton lifetime can consistently exceed the stringent experimental lower bound of $5.9 \times 10^{33}$ years established by the Super-Kamiokande collaboration.

\thispagestyle{empty}

\newpage
\renewcommand{\thefootnote}{\arabic{footnote}}
\setcounter{footnote}{0}
\baselineskip 18pt

\section{Introduction}

GUTs, 
 particularly those based on the $SU(5)$ gauge group
 \cite{{su5GUT}}, 
 provide an elegant and predictive framework for unifying the SM gauge interactions.
The supersymmetric extension of the $SU(5)$ model is especially attractive,
 as it successfully achieves precise gauge coupling unification at a high scale and offers a resolution to the hierarchy problem 
 \cite{SoftBrokenSUSY-SU5,Sakai-NaturalnessGUTS}.
Despite these theoretical triumphs, 
 the minimal supersymmetric $SU(5)$ GUT faces a critical phenomenological hurdle: 
 the problem of rapid proton decay
 \cite{Sakai:1981pk,Weinberg:1981wj}. 
The exchange of colored Higgsinos generates dimension-five effective operators that typically lead to a proton lifetime far shorter than allowed by experimental observations. 
Most notably, the stringent bound of $\tau(p \to K^+ \bar{\nu}) > 5.9 \times 10^{33}$ years, 
 set by the Super-Kamiokande collaboration \cite{Super-K-Limit}, 
 strongly restricts the model's viability.

Remedying the dimension-five proton decay problem typically requires modifications of the GUT-scale boundary conditions.
Typical solutions include introducing higher-dimensional representations to the Higgs sector (such as a $\mathbf{45}$-dimensional field) or incorporating non-renormalizable interactions
 \cite{Georgi:1979df,Giveon:1991zm}.
However,
 these extensions inevitably introduce a large number of additional,
 unconstrained parameters in the Yukawa matrices and the potential sector.
As a consequence,
 confronting such models with stringent experimental constraints becomes technically demanding due to the so-called ``curse of dimensionality.''
In particular,
 a conventional brute-force scan over the resulting high-dimensional parameter space—spanned by numerous mixing angles, CP phases, and mass scales—is computationally prohibitive.

In recent years, 
 machine learning techniques have emerged as an indispensable tool in high-energy physics
 \cite{BraneswithLearning,FlavorwithLearning,TruthBeautywithLearning}. 
They have proven exceptionally effective in exploring vast theoretical landscapes and optimizing complex models where traditional grid-search methods fail \cite{TensorFlow, Adam}. 
Inspired by recent successful applications of machine learning algorithms to resolve the fermion mass problem in GUTs,
 this paper adopts a similar computational approach to tackle the dimension-five proton decay problem.

We formulate our search for realistic,
 safe parameter configurations as an optimization problem.
By defining a loss function directly tied to the theoretical prediction of the $p \to K^+ \bar{\nu}$ decay rate,
 we employ the Adam optimization algorithm to navigate the 33-dimensional parameter space of a modified supersymmetric $SU(5)$ model.
We aim to identify regions in the parameter space that are highly optimized to suppress the decay width while remaining fully consistent with low-energy SM observables.

The remainder of this paper is organized as follows:
 Section \ref{section-model} details the structure of the supersymmetric $SU(5)$ model and the mathematical expressions for the dimension-five operators contributing to proton decay.
In Section \ref{section-Numerical-AnalysisOnYukawa},
 we establish our numerical setup, the boundary conditions,
 and the definition of the loss function. 
Section \ref{section-Numerical-Results} presents the results of the machine learning optimization across various values of $\tan \beta$.
Finally, Section \ref{section-summary} provides a summary of our findings and concluding remarks.

\section{Model}
\label{section-model}

\subsection{Model Definition}

The model we consider is a SUSY $SU(5)$ GUT model that contains three generations of matter fields in ${\bf \ov{5}}+{\bf 10}$ representations,
 electroweak-symmetry-breaking Higgs fields in ${\bf 5}+{\bf \ov{5}}+{\bf 45}+{\bf \ov{45}}$ representations,
 and a GUT-breaking field in ${\bf 24}$ representation.
$R$-parity $-1$ is assigned to the matter fields, and $+1$ to the rest.
The superpotential reads
\ba
W \ &=\ (Y_5)_{ij} \, {\bf 10}^i {\bf 10}^j {\bf 5}_H + (Y_{45})_{ij} \, {\bf 10}^i {\bf 10}^j {\bf 45}_H 
+ (Y_{\ov{5}})_{ij} \, {\bf 10}^i {\bf \ov{5}}^j\, {\bf \ov{5}}_H + (Y_{\ov{45}})_{ij} \, {\bf 10}^i {\bf \ov{5}}^j\, {\bf \ov{45}}_H
\nn\\
&+m \, {\bf 5}_H{\bf \ov{5}}_H + M \, {\bf 45}_H{\bf \ov{45}}_H 
+ \lambda \, {\bf 24}_H {\bf 5}_H{\bf \ov{5}}_H + \kappa \, {\bf 24}_H {\bf 5}_H{\bf \ov{45}}_H + \kappa' \, {\bf 24}_H {\bf \ov{5}}_H {\bf 45}_H 
\nn\\
&+ \lambda_a \, ({\bf 24}_H {\bf 45}_H{\bf \ov{45}}_H)_a + \lambda_b \, ({\bf 24}_H {\bf 45}_H{\bf \ov{45}}_H)_b,
\label{superpotential}
\end{align}
 where ${\bf \ov{5}}^i,{\bf 10}^i$ denote the matter fields with $i$ being the flavor index, 
 ${\bf 5}_H,{\bf \ov{5}}_H,{\bf 45}_H,{\bf \ov{45}}_H$ denote the electroweak-symmetry-breaking Higgs fields,
 and ${\bf 24}_H$ denotes the GUT-breaking field.
$Y_5$ has symmetric flavor indices and $Y_{45}$ has antisymmetric ones, i.e., $(Y_5)_{ij}=(Y_5)_{ji}$ and $(Y_{45})_{ij}=-(Y_{45})_{ji}$.
The two different couplings of ${\bf 45}_H,{\bf \ov{45}}_H,{\bf 24}_H$ are labelled by $a,b$.

The $({\bf 1},{\bf 2},\pm\frac{1}{2})$ components of ${\bf 5}_H,{\bf \ov{5}}_H,{\bf 45}_H,{\bf \ov{45}}_H$ in $SU(3)_C\times SU(2)_L\times U(1)_Y$ SM gauge group give the MSSM Higgs fields.
The $({\bf 3},{\bf 1},-\frac{1}{3})+({\bf \ov{3}},{\bf 1},\frac{1}{3})$ components of ${\bf 5}_H,{\bf \ov{5}}_H,{\bf 45}_H,{\bf \ov{45}}_H$,
 and the $({\bf 3},{\bf 3},-\frac{1}{3})+({\bf \ov{3}},{\bf 3},\frac{1}{3})$ and $({\bf \ov{3}},{\bf 1},\frac{4}{3})+({\bf 3},{\bf 1},-\frac{4}{3})$ components of ${\bf 45}_H,{\bf \ov{45}}_H$,
 which we call colored Higgs fields,
 give rise to rapid colored-Higgsino-mediated proton decay.
\\

We write the $({\bf 1},{\bf 2},\pm\frac{1}{2})$ components of ${\bf 5}_H,{\bf \ov{5}}_H$ as $({\bf 1},{\bf 2},\pm\frac{1}{2})_{5H}$,
 and those of ${\bf 45}_H,{\bf \ov{45}}_H$ as $({\bf 1},{\bf 2},\pm\frac{1}{2})_{45H}$.
We write the $({\bf 3},{\bf 1},-\frac{1}{3})+({\bf \ov{3}},{\bf 1},\frac{1}{3})$ components of ${\bf 5}_H,{\bf \ov{5}}_H$ as $({\bf 3},{\bf 1},-\frac{1}{3})_{5H}+({\bf \ov{3}},{\bf 1},\frac{1}{3})_{5H}$,
 and those of ${\bf 45}_H,{\bf \ov{45}}_H$ as $({\bf 3},{\bf 1},-\frac{1}{3})_{45H}+({\bf \ov{3}},{\bf 1},\frac{1}{3})_{45H}$.
We write the $({\bf 3},{\bf 3},-\frac{1}{3})+({\bf \ov{3}},{\bf 3},\frac{1}{3})$ and $({\bf \ov{3}},{\bf 1},\frac{4}{3})+({\bf 3},{\bf 1},-\frac{4}{3})$ components of 
 ${\bf 45}_H,{\bf \ov{45}}_H$ as $({\bf 3},{\bf 3},-\frac{1}{3})_H+({\bf \ov{3}},{\bf 3},\frac{1}{3})_H$ and $({\bf \ov{3}},{\bf 1},\frac{4}{3})_H+({\bf 3},{\bf 1},-\frac{4}{3})_H$, respectively.
We denote the isospin-doublet quark, isospin-singlet up-type quark, isospin-singlet down-type quark, isospin-doublet lepton, and charged lepton fields of MSSM by $Q^i,U^{c\,i},D^{c\,i},L^i,E^{c\,i}$, respectively.
\\

${\bf 24}_H$ gains a VEV as ${\bf 24}_H=\dfrac{v}{\sqrt{30}}$diag$(2,2,2,-3,-3)$ and breaks $SU(5)$ into $SU(3)_C\times SU(2)_L\times U(1)_Y$.
Then, the Yukawa couplings, the mass term involving the Higgs doublets, and the mass terms involving the colored Higgs fields are given by
\ba
W &\supset W_{\rm Yukawa1} + W_{\rm Yukawa2} + W_{\rm mass1} + W_{\rm mass2} \ ,
\end{align}
 where
 \ba
W_{\rm Yukawa1} = &-2(Y_5)_{ij}\,Q^iU^{c\,j}({\bf 1},{\bf 2},\tfrac{1}{2})_{5H} - \frac{1}{\sqrt{2}}(Y_{\ov{5}})_{ij}\,Q^iD^{c\,j}({\bf 1},{\bf 2},-\tfrac{1}{2})_{5H} 
- \frac{1}{\sqrt{2}}(Y_{\ov{5}})_{ij}\,E^{c\,i}L^j({\bf 1},{\bf 2},-\tfrac{1}{2})_{5H} 
\nn\\
&-\sqrt{\frac{2}{3}}(Y_{45})_{ij}\,Q^iU^{c\,j}({\bf 1},{\bf 2},\tfrac{1}{2})_{45H} - \frac{1}{2\sqrt{3}}(Y_{\ov{45}})_{ij}\,Q^iD^{c\,j}({\bf 1},{\bf 2},-\tfrac{1}{2})_{45H} 
\nn\\
&+ \frac{\sqrt{3}}{2}(Y_{\ov{45}})_{ij}\,E^{c\,i}L^j({\bf 1},{\bf 2},-\tfrac{1}{2})_{45H},
\\
W_{\rm Yukawa2} =&-2(Y_5)_{ij}\,Q^iQ^j({\bf 3},{\bf 1},-\tfrac{1}{3})_{5H} -2(Y_5)_{ij}\,E^{c\,i}U^{c\,j}({\bf 3},{\bf 1},-\tfrac{1}{3})_{5H},
\nn\\
&+\frac{1}{\sqrt{2}}(Y_{\ov{5}})_{ij}\,Q^iL^j({\bf \ov{3}},{\bf 1},\tfrac{1}{3})_{5H} +\frac{1}{\sqrt{2}}(Y_{\ov{5}})_{ij}\,U^{c\,i}D^{c\,j}({\bf \ov{3}},{\bf 1},\tfrac{1}{3})_{5H}
\nn\\
&-\sqrt{2}(Y_{45})_{ij}\,E^{c\,i}U^{c\,j}({\bf 3},{\bf 1},-\tfrac{1}{3})_{45H}
\nn\\
&-\frac{1}{2}(Y_{\ov{45}})_{ij}\,Q^iL^j({\bf \ov{3}},{\bf 1},\tfrac{1}{3})_{45H} +\frac{1}{2}(Y_{\ov{45}})_{ij}\,U^{c\,i}D^{c\,j}({\bf \ov{3}},{\bf 1},\tfrac{1}{3})_{45H}
\nn\\
&-\sqrt{2}(Y_{45})_{ij}\,Q^iQ^j({\bf 3},{\bf 3},-\tfrac{1}{3})_{H} + \sqrt{2}(Y_{45})_{ij}\,U^{c\,i}U^{c\,j}({\bf \ov{3}},{\bf 1},\tfrac{4}{3})_{H}
\nn\\
&+(Y_{\ov{45}})_{ij}\,Q^iL^j({\bf \ov{3}},{\bf 3},\tfrac{1}{3})_{H} - (Y_{\ov{45}})_{ij}\,E^{c\,i}D^{c\,j}({\bf 3},{\bf 1},-\tfrac{4}{3})_{H},
\\
 W_{\rm mass1} = &\begin{pmatrix}
      ({\bf 1},{\bf 2},\tfrac{1}{2})_{5H} & ({\bf 1},{\bf 2},\tfrac{1}{2})_{45H} \\
   \end{pmatrix}  
{\cal M}_{\rm doublet}
   \begin{pmatrix}
      ({\bf 1},{\bf 2},-\tfrac{1}{2})_{5H} \\
      ({\bf 1},{\bf 2},-\tfrac{1}{2})_{45H} \\
   \end{pmatrix},
\\
W_{\rm mass2}= &\begin{pmatrix}
      ({\bf 3},{\bf 1},-\tfrac{1}{3})_{5H} & ({\bf 3},{\bf 1},-\tfrac{1}{3})_{45H} \\
   \end{pmatrix}  
{\cal M}_{\rm triplet}
   \begin{pmatrix}
      ({\bf \ov{3}},{\bf 1},\tfrac{1}{3})_{5H} \\
      ({\bf \ov{3}},{\bf 1},\tfrac{1}{3})_{45H} \\
   \end{pmatrix}
\nn\\
&+\left(M-\frac{1}{2\sqrt{30}}(6\lambda_a+\lambda_b)v\right)({\bf 3},{\bf 3},-\tfrac{1}{3})_{H}({\bf \ov{3}},{\bf 3},\tfrac{1}{3})_{H}
\nn\\
&+\left(M-\frac{1}{\sqrt{30}}(-2\lambda_a+3\lambda_b)v\right)({\bf \ov{3}},{\bf 1},\tfrac{4}{3})_{H}({\bf 3},{\bf 1},-\tfrac{4}{3})_{H}
\end{align} 
with
\begin{align}
{\cal M}_{\rm doublet} &= \begin{pmatrix}
      m-\frac{3}{\sqrt{30}}\lambda v & -\frac{\sqrt{5}}{4}\kappa v \\
      -\frac{\sqrt{5}}{4}\kappa' v & M-\frac{1}{8\sqrt{30}}(14\lambda_a+19\lambda_b)v \\
   \end{pmatrix},
 \label{mass1}
 \\
{\cal M}_{\rm triplet} &= \begin{pmatrix}
      m+\frac{2}{\sqrt{30}}\lambda v & -\frac{\sqrt{5}}{2\sqrt{3}}\kappa v \\
      -\frac{\sqrt{5}}{2\sqrt{3}}\,\kappa' v & M-\frac{1}{4\sqrt{30}}(2\lambda_a-3\lambda_b)v \\
   \end{pmatrix}.
    \label{mass2}
\end{align}
\\

\subsection{Expressions of the GUT Yukawa couplings}

We fine-tune $m,M$ so that the mass matrix ${\cal M}_{\rm doublet}$ in Eq.~(\ref{mass1}) has one zero singular value,
 corresponding to the $\mu$-term of MSSM (which should be much smaller than the GUT scale).
Let us write the singular value decomposition of ${\cal M}_{\rm doublet}$ as
\ba
 \begin{pmatrix} 
         U_{11} & U_{12} \\
         U_{21} & U_{22} \\
      \end{pmatrix}
{\cal M}_{\rm doublet}
   \begin{pmatrix} 
         V_{11} & V_{12} \\
         V_{21} & V_{22} \\
      \end{pmatrix}
       \ = \   
      \begin{pmatrix} 
         0 & 0 \\
         0 & M_D \\
      \end{pmatrix},
\end{align}
 where $\begin{pmatrix} 
         U_{11} & U_{12} \\
         U_{21} & U_{22} \\
      \end{pmatrix}$, $\begin{pmatrix} 
         V_{11} & V_{12} \\
         V_{21} & V_{22} \\
      \end{pmatrix}$
      are unitary matrices, and $M_D$ is a GUT-scale mass.
Then the components of $({\bf 1},{\bf 2},\tfrac{1}{2})_{5H}$, $({\bf 1},{\bf 2},\tfrac{1}{2})_{45H}$, $({\bf 1},{\bf 2},-\tfrac{1}{2})_{5H}$, $({\bf 1},{\bf 2},-\tfrac{1}{2})_{45H}$
 that correspond to the MSSM Higgs fields, $H_u,H_d$, are found to be
\begin{align}
({\bf 1},{\bf 2},\tfrac{1}{2})_{5H} \supset U_{11}H_u, \ \ ({\bf 1},{\bf 2},\tfrac{1}{2})_{45H} \supset U_{12}H_u, \ \ ({\bf 1},{\bf 2},-\tfrac{1}{2})_{5H} \supset V_{11}H_d,
\ \ ({\bf 1},{\bf 2},-\tfrac{1}{2})_{45H} \supset V_{21}H_d.
\end{align}
Accordingly, the up-type quark, down-type quark and charged lepton Yukawa coupling matrices of MSSM, $Y_u,Y_d,Y_e$, are matched to $Y_5,Y_{45},Y_{\ov{5}},Y_{\ov{45}}$ at the GUT scale ($\simeq 2\times10^{16}$~GeV) as
\footnote{
The MSSM Yukawa coupling matrices are defined as
\ba
W_{\rm MSSM} \supset (Y_u)_{ij}\,Q^i U^{c\,j} H_u + (Y_d)_{ij}\,Q^i D^{c\,j} H_d + (Y_e)_{ij}\,L^i E^{c\,j} H_d.
\end{align}
}
\ba
Y_u = -2U_{11} \, Y_5 - \sqrt{\frac{2}{3}}U_{12} \, Y_{45}, \ \ \ \ Y_d = -\frac{1}{\sqrt{2}}V_{11} \, Y_{\ov{5}} - \frac{1}{2\sqrt{3}}V_{21} \, Y_{\ov{45}},
\ \ \ \ Y_e = -\frac{1}{\sqrt{2}}V_{11} \, Y_{\ov{5}}^T + \frac{\sqrt{3}}{2}V_{21} \, Y_{\ov{45}}^T.
\label{mssmyukawa}
\end{align}

In the flavor bases of ${\bf 10}^i$ and ${\bf \ov{5}}^i$ where the down-type quark Yukawa coupling matrix $Y_d$ is diagonal,
 the MSSM Yukawa coupling matrices can be expressed as
\ba
Y_u = V_{CKM}^T   \begin{pmatrix} 
      y_u & 0 & 0 \\
      0 & y_c & 0 \\
      0 & 0 & y_t \\
   \end{pmatrix} U_u,
\ \ \ \ \ 
Y_d = \begin{pmatrix} 
      y_d & 0 & 0 \\
      0 & y_s & 0 \\
      0 & 0 & y_b \\
   \end{pmatrix},
\ \ \ \ \ 
Y_e = V_e\begin{pmatrix} 
      y_e & 0 & 0 \\
      0 & y_\mu & 0 \\
      0 & 0 & y_\tau \\
   \end{pmatrix}U_e,
\label{yukawasindldrbasis}
\end{align}
 where $V_{CKM}$ is the CKM matrix,
 $y_u,y_c,y_t,y_d,y_s,y_b,y_e,y_\mu,y_\tau$ are the singular values of the Yukawa coupling matrices,
 and $U_u,U_e,V_e$ are undetermined unitary matrices.
From Eq.~(\ref{mssmyukawa}) and symmetry of $Y_5$ and antisymmetry of $Y_{45}$, Yukawa coupling matrices $Y_5,Y_{\ov{5}},Y_{45},Y_{\ov{45}}$ in the same flavor bases are obtained as
\ba
Y_5 &= -\frac{1}{4U_{11}}\left[V_{CKM}^T   \begin{pmatrix} 
      y_u & 0 & 0 \\
      0 & y_c & 0 \\
      0 & 0 & y_t \\
   \end{pmatrix} U_u + U_u^T   \begin{pmatrix} 
      y_u & 0 & 0 \\
      0 & y_c & 0 \\
      0 & 0 & y_t \\
   \end{pmatrix} V_{CKM}\right],
\label{y5}\\
Y_{45} &= -\frac{\sqrt{3}}{2\sqrt{2}U_{12}}\left[V_{CKM}^T   \begin{pmatrix} 
      y_u & 0 & 0 \\
      0 & y_c & 0 \\
      0 & 0 & y_t \\
   \end{pmatrix} U_u - U_u^T   \begin{pmatrix} 
      y_u & 0 & 0 \\
      0 & y_c & 0 \\
      0 & 0 & y_t \\
   \end{pmatrix} V_{CKM}\right],
 \label{y45}\\
 Y_{\ov{5}} &= -\frac{1}{2\sqrt{2}V_{11}}\left[
 3\begin{pmatrix} 
      y_d & 0 & 0 \\
      0 & y_s & 0 \\
      0 & 0 & y_b \\
   \end{pmatrix}
   +
   U_e^T\begin{pmatrix} 
      y_e & 0 & 0 \\
      0 & y_\mu & 0 \\
      0 & 0 & y_\tau \\
   \end{pmatrix}V_e^T
   \right],
\label{y5bar}\\
Y_{\ov{45}} &= -\frac{\sqrt{3}}{2V_{21}}\left[
\begin{pmatrix} 
      y_d & 0 & 0 \\
      0 & y_s & 0 \\
      0 & 0 & y_b \\
   \end{pmatrix}
   -
   U_e^T\begin{pmatrix} 
      y_e & 0 & 0 \\
      0 & y_\mu & 0 \\
      0 & 0 & y_\tau \\
   \end{pmatrix}V_e^T
   \right],
   \label{y45bar}
\end{align}
 where $V_{CKM}$ and $y_u,y_c,y_t,y_d,y_s,y_b,y_e,y_\mu,y_\tau$ are evaluated at the GUT scale.
\\

\subsection{Dimension-five operators contributing to proton decay}

Integrating out the colored Higgs fields, we obtain dimension-five operators contributing to proton decay
\ba
W_5 = (C_{5L})_{ijkl} (Q^iQ^j)(Q^kL^l) + (C_{5R})_{ijkl} E^{c\,i}U^{c\,j}U^{c\,k}D^{c\,l},
\end{align}
 where isospin indices are summed in each bracket in the first term.
The Wilson coefficients $C_{5L},C_{5R}$ at the GUT scale are expressed with the GUT Yukawa couplings as
\ba
(C_{5L})_{ijkl} =    &-2\begin{pmatrix} 
      \frac{1}{\sqrt{2}}(Y_{\ov{5}})_{kl} & -\frac{1}{2}(Y_{\ov{45}})_{kl}
   \end{pmatrix}{\cal M}_{\rm triplet}^{-1}
   \begin{pmatrix} 
      -2(Y_5)_{ij} \\
      0
   \end{pmatrix}
+ \sqrt{2}\frac{1}{2}\frac{ (Y_{\ov{45}})_{il}(Y_{45})_{kj} + (Y_{\ov{45}})_{jl}(Y_{45})_{ki} }{M-\frac{1}{2\sqrt{30}}(6\lambda_a+\lambda_b)v},
\label{c5l}\\
(C_{5R})_{ijkl} =    &-\begin{pmatrix} 
      \frac{1}{\sqrt{2}}(Y_{\ov{5}})_{kl} & \frac{1}{2}(Y_{\ov{45}})_{kl}
   \end{pmatrix}{\cal M}_{\rm triplet}^{-1}
   \begin{pmatrix} 
      -2(Y_5)_{ij} \\
      -\sqrt{2}(Y_{45})_{ij}
   \end{pmatrix}
   +\begin{pmatrix} 
      \frac{1}{\sqrt{2}}(Y_{\ov{5}})_{jl} & \frac{1}{2}(Y_{\ov{45}})_{jl}
   \end{pmatrix}{\cal M}_{\rm triplet}^{-1}
   \begin{pmatrix} 
      -2(Y_5)_{ik} \\
      -\sqrt{2}(Y_{45})_{ik}
   \end{pmatrix} 
\nn\\   
&+ 2\sqrt{2}\frac{(Y_{\ov{45}})_{il}(Y_{45})_{jk}}{M-\frac{1}{\sqrt{30}}(-2\lambda_a+3\lambda_b)v}.
\label{c5r}
\end{align}
\\

\subsection{Calculation of the $p\to K^+ \bar{\nu}$ partial width}

We concentrate on the $p\to K^+ \bar{\nu}$ mode, since it tends to have large branching ratio and it is also experimentally severely constrained.
The $p\to K^+ \bar{\nu}$ partial width is the sum of the $p\to K^+ \bar{\nu}_\tau$, $K^+ \bar{\nu}_\mu$, $K^+ \bar{\nu}_e$ partial widths, which are given by
\footnote{
In the Wilson coefficients, for example, flavor index $..._{u_L}$ means that it is associated with a matter field in the same {\bf 10} representation field as the left-handed up quark $u_L$.
}

\ba
&\Gamma(p\to K^+\bar{\nu}_\tau)
 = 
\nn\\
&\frac{m_N}{64\pi}\left(1-\frac{m_K^2}{m_N^2}\right)^2
\left\vert \beta_H(\mu_{\rm had})\frac{1}{f_\pi}\left\{
\left(1+\frac{D}{3}+F\right)(C_{LL})_{u_L d_L s_L \tau_L}(\mu_{\rm had})+\frac{2D}{3}(C_{LL})_{u_L s_L d_L \tau_L}(\mu_{\rm had})\right\}
\right.
\nn\\
&\left. \ \ \ \ \ \ \ \ \ \ \ \ +\alpha_H(\mu_{\rm had})\frac{1}{f_\pi}\left\{
\left(1+\frac{D}{3}+F\right)(C_{LR})_{\tau_L s_L u_R d_R}(\mu_{\rm had})+\frac{2D}{3}(C_{LR})_{\tau_L d_L u_R s_R}(\mu_{\rm had})
\right\}\right\vert^2,
\label{nutau}
\\
&\Gamma(p\to K^+\bar{\nu}_{\ell=e,\mu})
= 
\nn\\&\frac{m_N}{64\pi}\left(1-\frac{m_K^2}{m_N^2}\right)^2
\left\vert \beta_H(\mu_{\rm had})\frac{1}{f_\pi}\left\{
\left(1+\frac{D}{3}+F\right)(C_{LL})_{u_L d_L s_L \ell_L}(\mu_{\rm had})+\frac{2D}{3}(C_{LL})_{u_L s_L d_L \ell_L}(\mu_{\rm had})\right\}
\right\vert^2,
\label{numue}
\end{align} 
 where $\mu_{\rm had}$ denotes the hadronic scale, $\alpha_H,\beta_H$ hadronic matrix elements, and $D,F$ parameters of the baryon chiral Lagrangian.
Here $C_{RL},C_{LL}$ are the Wilson coefficients of dimension-six operators defined as
 $-{\cal L}_6 = (C_{LL})_{ijkl}(\psi_{u_L}^i\psi_{d_L}^j)(\psi_{d_L}^k\psi_{\nu_L}^l)+(C_{LR})_{ijkl}(\psi_{\nu_L}^i\psi_{d_L}^j)(\psi_{u_R^c}^k\psi_{d_R^c}^l)$
 with $\psi$ denoting SM left-handed Weyl spinors whose spinor indices are summed in each bracket.
They satisfy
\ba
&(C_{LL})_{u_L d_L s_L \ell_L}(\mu_{\rm had}) = A_{LL}(\mu_{\rm had},\mu_{\rm SUSY})
\frac{M_{\widetilde{W}}}{m_{\tilde{q}}^2}
\frac{{\cal F}}{16\pi^2}\ g_2^2 
\nn\\
&\ \ \ \ \ \ \ \ \ \ \ \ \ \ \ \ \ \ \ \ \ \ \ \ \ \times A_L(\mu_{\rm SUSY},\mu_{\rm GUT}) \left\{ (C_{5L})_{u_L d_L s_L \ell_L}(\mu_{\rm GUT})-(C_{5L})_{d_L s_L u_L \ell_L}(\mu_{\rm GUT})\right\},
\label{cll1}
\\
&(C_{LL})_{u_L d_L s_L \tau_L}(\mu_{\rm had}) = A_{LL}(\mu_{\rm had},\mu_{\rm SUSY})
\frac{M_{\widetilde{W}}}{m_{\tilde{q}}^2}
\frac{{\cal F}_\tau}{16\pi^2}\ g_2^2
\nn\\
&\ \ \ \ \ \ \ \ \ \ \ \ \ \ \ \ \ \ \ \ \ \ \ \ \ \ \times A_L^\tau(\mu_{\rm SUSY},\mu_{\rm GUT}) \left\{ (C_{5L})_{u_L d_L s_L \tau_L}(\mu_{\rm GUT})-(C_{5L})_{d_L s_L u_L \tau_L}(\mu_{\rm GUT})\right\},
\label{cll2}
\\
&(C_{LL})_{u_L s_L d_L \ell_L}(\mu_{\rm had}) = A_{LL}(\mu_{\rm had},\mu_{\rm SUSY})
\frac{M_{\widetilde{W}}}{m_{\tilde{q}}^2}
\frac{{\cal F}}{16\pi^2}\ g_2^2 
\nn\\
&\ \ \ \ \ \ \ \ \ \ \ \ \ \ \ \ \ \ \ \ \ \ \ \ \ \times A_L(\mu_{\rm SUSY},\mu_{\rm GUT}) \left\{ (C_{5L})_{u_L s_L d_L \ell_L}(\mu_{\rm GUT})-(C_{5L})_{d_L s_L u_L \ell_L}(\mu_{\rm GUT})\right\},
\label{cll3}
\\
&(C_{LL})_{u_L s_L d_L \tau_L}(\mu_{\rm had}) = A_{LL}(\mu_{\rm had},\mu_{\rm SUSY})
\frac{M_{\widetilde{W}}}{m_{\tilde{q}}^2}
\frac{{\cal F}_\tau}{16\pi^2}\ g_2^2
\nn\\
&\ \ \ \ \ \ \ \ \ \ \ \ \ \ \ \ \ \ \ \ \ \ \ \ \ \ \times A_L^\tau(\mu_{\rm SUSY},\mu_{\rm GUT}) \left\{ (C_{5L})_{u_L s_L d_L \tau_L}(\mu_{\rm GUT})-(C_{5L})_{d_L s_L u_L \tau_L}(\mu_{\rm GUT})\right\},
\label{cll4}
\\
&(C_{LR})_{\tau_L s_L u_R d_R}(\mu_{\rm had}) = A_{LR}(\mu_{\rm had},\mu_{\rm SUSY})
\frac{\mu_H}{m_{\tilde{t}_R}^2}
\frac{{\cal F}'}{16\pi^2}(V_{CKM}^*)_{t_Ls_L}\ y_t y_\tau
\nn\\
&\ \ \ \ \ \ \ \ \ \ \ \ \ \ \ \ \ \ \ \ \ \ \ \ \ \ \ \ \ \ \ \ \ \ \ \ \ \ \ \ \ \ \ \ \ \  \ \ \ \ \ \ \ \ \times A_R^{\tau t}(\mu_{\rm SUSY},\mu_{\rm GUT})(C_{5R})_{\tau_R t_R u_R d_R}(\mu_{\rm GUT}),
\label{crl1}
\\
&(C_{LR})_{\tau_L d_L u_R s_R}(\mu_{\rm had}) = A_{LR}(\mu_{\rm had},\mu_{\rm SUSY})
\frac{\mu_H}{m_{\tilde{t}_R}^2}
\frac{{\cal F}'}{16\pi^2}(V_{CKM}^*)_{t_Ld_L}\ y_t y_\tau
\nn\\
&\ \ \ \ \ \ \ \ \ \ \ \ \ \ \ \ \ \ \ \ \ \ \ \ \ \ \ \ \ \ \ \ \ \ \ \ \ \ \ \ \ \ \ \ \ \  \ \ \ \ \ \ \ \ \times 
A_R^{\tau t}(\mu_{\rm SUSY},\mu_{\rm GUT})(C_{5R})_{\tau_R t_R u_R s_R}(\mu_{\rm GUT}),
\label{crl2}
\end{align}
 where $\mu_{\rm SUSY}$ denotes the soft SUSY breaking scale and $\mu_{\rm GUT}$ the GUT scale.
Here ${\cal F},{\cal F}_\tau,{\cal F}'$ are loop functions defined as
 ${\cal F} =\frac{1}{z-w}(\frac{z}{1-z}\log z - \frac{w}{1-w}\log w) + \frac{1}{z-1}(\frac{z}{1-z}\log z+1)$, ${\cal F}_\tau =\frac{1}{z-w_\tau}(\frac{z}{1-z}\log z - \frac{w_\tau}{1-w_\tau}\log w_\tau) + \frac{1}{z-1}(\frac{z}{1-z}\log z+1)$, ${\cal F}' =\frac{1}{x-y}(\frac{x}{1-x}\log x - \frac{y}{1-y}\log y)$
 with $x=|\mu_H|^2/m_{\tilde{t}_R}^2$, $y=m_{\tilde{\tau}_R}^2/m_{\tilde{t}_R}^2$, $z=|M_{\widetilde{W}}|^2/m_{\tilde{q}}^2$, $w=m_{\tilde{\ell}_L}^2/m_{\tilde{q}}^2$, $w_\tau=m_{\tilde{\tau}_L}^2/m_{\tilde{q}}^2$,
 and 
$\mu_H,m_{\tilde{t}_R},m_{\tilde{\tau}_R},$
$M_{\widetilde{W}},m_{\tilde{\ell}_L},m_{\tilde{q}_L},m_{\tilde{\tau}_L}$ denoting the pole masses of Higgsinos, isospin-singlet top squark, isospin-singlet stau,
 Winos, isospin-doublet smuon/selectron, 1st/2nd generation isospin-doublet squarks, and isospin-doublet stau, respectively.
$(V_{CKM}^*)_{t_Ls_L},(V_{CKM}^*)_{t_Ld_L}$ are components of the CKM matrix and $y_t,y_\tau,g_2$ are respectively the top quark Yukawa coupling, tau lepton Yukawa coupling and weak gauge coupling,
 all of which are evaluated in MSSM at scale $\mu=\mu_{\rm SUSY}$.
Finally, $A_{LL}(\mu_{\rm had},\mu_{\rm SUSY})$ accounts for RG evolutions of Wilson coefficients $(C_{LL})_{u_L d_L s_L \tau_L},$
$(C_{LL})_{u_L d_L s_L \ell_L}$ in SM from scale $\mu_{\rm SUSY}$ to $\mu_{\rm had}$, and $A_{LR}(\mu_{\rm had},\mu_{\rm SUSY})$ those of Wilson coefficients $(C_{LR})_{\tau_L s_L u_R d_R},$
$(C_{LR})_{\tau_L d_L u_R s_R}$.
Also, $A_L(\mu_{\rm SUSY},\mu_{\rm GUT})$ accounts for RG evolutions of Wilson coefficients $(C_{5L})_{u_L d_L s_L \ell_L},$
$(C_{5L})_{d_L s_L u_L \ell_L}$ $(\ell=e,\mu)$ from scale $\mu_{\rm GUT}$ to $\mu_{\rm SUSY}$,
 $A_L^\tau(\mu_{\rm SUSY},\mu_{\rm GUT})$ those of Wilson coefficients $(C_{5L})_{u_L d_L s_L \tau_L},(C_{5L})_{d_L s_L u_L \tau_L}$,
 and $A_R^{\tau t}(\mu_{\rm SUSY},\mu_{\rm GUT})$ those of Wilson coefficients $(C_{5R})_{\tau_R t_R u_R d_R},$
 $(C_{5R})_{\tau_R t_R u_R s_R}$.

The Wilson coefficients on the right-hand side of Eqs.~(\ref{cll1})-(\ref{crl2}) are obtained from Eqs.~(\ref{c5l}),(\ref{c5r}). 
The Yukawa coupling components they contain are re-expressed with Eqs.~(\ref{y5})-(\ref{y45bar}) as
\ba
(Y_5)_{u_Ld_L}&= -\frac{1}{4U_{11}}\left[\begin{pmatrix} 
      y_u & 0 & 0 \\
      0 & y_c & 0 \\
      0 & 0 & y_t \\
   \end{pmatrix} U_u + V_{CKM}^*U_u^T   \begin{pmatrix} 
      y_u & 0 & 0 \\
      0 & y_c & 0 \\
      0 & 0 & y_t \\
   \end{pmatrix} V_{CKM}\right]_{11},
\nn\\
(Y_5)_{u_Ls_L}&= [{\rm the \ same \ matrix \ as \ above}]_{12},
\nn\\
(Y_5)_{d_Ls_L}&= -\frac{1}{4U_{11}}\left[
V_{CKM}^T   \begin{pmatrix} 
      y_u & 0 & 0 \\
      0 & y_c & 0 \\
      0 & 0 & y_t \\
   \end{pmatrix} U_u + U_u^T   \begin{pmatrix} 
      y_u & 0 & 0 \\
      0 & y_c & 0 \\
      0 & 0 & y_t \\
   \end{pmatrix} V_{CKM}
\right]_{12},
\nn\\
(Y_{45})_{s_Ld_L}&= -(Y_{45})_{d_Ls_L} = -\frac{\sqrt{3}}{2\sqrt{2}U_{12}}\left[
V_{CKM}^T   \begin{pmatrix} 
      y_u & 0 & 0 \\
      0 & y_c & 0 \\
      0 & 0 & y_t \\
   \end{pmatrix} U_u - U_u^T   \begin{pmatrix} 
      y_u & 0 & 0 \\
      0 & y_c & 0 \\
      0 & 0 & y_t \\
   \end{pmatrix} V_{CKM}
   \right]_{21},
\nn\\
(Y_{45})_{d_Lu_L}&=-(Y_{45})_{u_Ld_L}= -\frac{\sqrt{3}}{2\sqrt{2}U_{12}}\left[
V_{CKM}^T\begin{pmatrix} 
      y_u & 0 & 0 \\
      0 & y_c & 0 \\
      0 & 0 & y_t \\
   \end{pmatrix} U_u V_{CKM}^\dagger - U_u^T   \begin{pmatrix} 
      y_u & 0 & 0 \\
      0 & y_c & 0 \\
      0 & 0 & y_t \\
   \end{pmatrix}
   \right]_{11},
\nn\\
(Y_{45})_{s_Lu_L}&=-(Y_{45})_{u_Ls_L}= [{\rm the \ same \ matrix \ as \ above}]_{21},
\nn\\
(Y_{\ov{5}})_{s_L\alpha_L}&= -\frac{1}{2\sqrt{2}V_{11}}\left[
 3\begin{pmatrix} 
      y_d & 0 & 0 \\
      0 & y_s & 0 \\
      0 & 0 & y_b \\
   \end{pmatrix}V_e^*
   +
   U_e^T\begin{pmatrix} 
      y_e & 0 & 0 \\
      0 & y_\mu & 0 \\
      0 & 0 & y_\tau \\
   \end{pmatrix}
   \right]_{2l},
\nn\\
(Y_{\ov{5}})_{d_L\alpha_L}&= [{\rm the \ same \ matrix \ as \ above}]_{1l},
\nn\\
(Y_{\ov{5}})_{u_L\alpha_L}&= -\frac{1}{2\sqrt{2}V_{11}}\left[
 3V_{CKM}^*\begin{pmatrix} 
      y_d & 0 & 0 \\
      0 & y_s & 0 \\
      0 & 0 & y_b \\
   \end{pmatrix}V_e^*
   +
   V_{CKM}^*U_e^T\begin{pmatrix} 
      y_e & 0 & 0 \\
      0 & y_\mu & 0 \\
      0 & 0 & y_\tau \\
   \end{pmatrix}
   \right]_{1l},
\nn\\
(Y_{\ov{45}})_{s_L\alpha_L}&= -\frac{\sqrt{3}}{2V_{21}}\left[
\begin{pmatrix} 
      y_d & 0 & 0 \\
      0 & y_s & 0 \\
      0 & 0 & y_b \\
   \end{pmatrix}V_e^*
   -
   U_e^T\begin{pmatrix} 
      y_e & 0 & 0 \\
      0 & y_\mu & 0 \\
      0 & 0 & y_\tau \\
   \end{pmatrix}
   \right]_{2l},
\nn\\
(Y_{\ov{45}})_{d_L\alpha_L}&= [{\rm the \ same \ matrix \ as \ above}]_{1l}, 
\nn\\
(Y_{\ov{45}})_{u_L\alpha_L}&= -\frac{\sqrt{3}}{2V_{21}}\left[
V_{CKM}^*\begin{pmatrix} 
      y_d & 0 & 0 \\
      0 & y_s & 0 \\
      0 & 0 & y_b \\
   \end{pmatrix}V_e^*
   -
   V_{CKM}^*U_e^T\begin{pmatrix} 
      y_e & 0 & 0 \\
      0 & y_\mu & 0 \\
      0 & 0 & y_\tau \\
   \end{pmatrix}
   \right]_{1l},
\label{YL}
\end{align}
where $l=1,2,3$ for $\alpha=e,\mu,\tau$,
and
\ba
(Y_5)_{\tau_Rt_R}&=-\frac{1}{4U_{11}}\left[U_e^*V_{CKM}^T   \begin{pmatrix} 
      y_u & 0 & 0 \\
      0 & y_c & 0 \\
      0 & 0 & y_t \\
   \end{pmatrix} + U_e^*U_u^T   \begin{pmatrix} 
      y_u & 0 & 0 \\
      0 & y_c & 0 \\
      0 & 0 & y_t \\
   \end{pmatrix} V_{CKM}U_u^\dagger \right]_{33},
\nn\\
(Y_5)_{\tau_Ru_R}&= [{\rm the \ same \ matrix \ as \ above}]_{31}
\nn\\
(Y_{45})_{\tau_Rt_R}&= -\frac{\sqrt{3}}{2\sqrt{2}U_{12}}\left[U_e^*V_{CKM}^T   \begin{pmatrix} 
      y_u & 0 & 0 \\
      0 & y_c & 0 \\
      0 & 0 & y_t \\
   \end{pmatrix} - U_e^*U_u^T   \begin{pmatrix} 
      y_u & 0 & 0 \\
      0 & y_c & 0 \\
      0 & 0 & y_t \\
   \end{pmatrix} V_{CKM}U_u^\dagger \right]_{33},
\nn\\
(Y_{45})_{\tau_Ru_R}&= [{\rm the \ same \ matrix \ as \ above}]_{31}
\nn\\
(Y_{45})_{t_Ru_R}&= -\frac{\sqrt{3}}{2\sqrt{2}U_{12}}\left[
U_u^*V_{CKM}^T   \begin{pmatrix} 
      y_u & 0 & 0 \\
      0 & y_c & 0 \\
      0 & 0 & y_t \\
   \end{pmatrix} - \begin{pmatrix} 
      y_u & 0 & 0 \\
      0 & y_c & 0 \\
      0 & 0 & y_t \\
   \end{pmatrix} V_{CKM}U_u^\dagger
 \right]_{31},
\nn\\
(Y_{\ov{5}})_{u_Rd_R}&=-\frac{1}{2\sqrt{2}V_{11}}\left[
 3U_u^*\begin{pmatrix} 
      y_d & 0 & 0 \\
      0 & y_s & 0 \\
      0 & 0 & y_b \\
   \end{pmatrix}
   +
   U_u^*U_e^T\begin{pmatrix} 
      y_e & 0 & 0 \\
      0 & y_\mu & 0 \\
      0 & 0 & y_\tau \\
   \end{pmatrix}V_e^T
   \right]_{11},
\nn\\
(Y_{\ov{5}})&_{u_Rs_R},\ (Y_{\ov{5}})_{t_Rd_R},\ (Y_{\ov{5}})_{t_Rs_R}= [{\rm the \ same \ matrix \ as \ above}]_{12}, \, _{31}, \, _{32},
\nn\\
(Y_{\ov{45}})_{u_Rd_R}&=-\frac{\sqrt{3}}{2V_{21}}\left[
U_u^*\begin{pmatrix} 
      y_d & 0 & 0 \\
      0 & y_s & 0 \\
      0 & 0 & y_b \\
   \end{pmatrix}
   -
   U_u^*U_e^T\begin{pmatrix} 
      y_e & 0 & 0 \\
      0 & y_\mu & 0 \\
      0 & 0 & y_\tau \\
   \end{pmatrix}V_e^T
   \right]_{11},
\nn\\
(Y_{\ov{45}})&_{u_Rs_R},\ (Y_{\ov{45}})_{t_Rd_R},\ (Y_{\ov{45}})_{t_Rs_R}= [{\rm the \ same \ matrix \ as \ above}]_{12}, \, _{31}, \, _{32},
\nn\\
(Y_{\ov{45}})_{\tau_Rd_R}&=-\frac{\sqrt{3}}{2V_{21}}\left[
U_e^*\begin{pmatrix} 
      y_d & 0 & 0 \\
      0 & y_s & 0 \\
      0 & 0 & y_b \\
   \end{pmatrix}
   -
   \begin{pmatrix} 
      y_e & 0 & 0 \\
      0 & y_\mu & 0 \\
      0 & 0 & y_\tau \\
   \end{pmatrix}V_e^T
   \right]_{31},
\nn\\
(Y_{\ov{45}})_{\tau_Rs_R}&= [{\rm the \ same \ matrix \ as \ above}]_{32}.
\label{YR}
\end{align}
\\

\section{Numerical Analysis on Yukawa Couplings}
\label{section-Numerical-AnalysisOnYukawa}

We vary the values of the undetermined unitary matrices $U_u,\,U_e,\,V_e$ in Eqs.~(\ref{YL}),(\ref{YR})
 and parameters $m,M,\lambda,\kappa,\kappa',\lambda^a,\lambda^b$ 
 in ${\cal M}_{\rm doublet}$ Eq.~(\ref{mass1}) and ${\cal M}_{\rm triplet}$ Eq.~(\ref{mass2}),
 under the condition that ${\cal M}_{\rm doublet}$ have one zero singular value (corresponding to the MSSM $\mu$-term).
Thereby we minimize the partial width of the $p\to K^+ \bar{\nu}$ decay, which is the sum of Eqs.~(\ref{nutau}),(\ref{numue}).

In our numerical analysis, the benchmark SUSY particle mass spectrum is as follows:
\begin{align}
 &\mu_H = 3000~{\rm GeV}, \ \ \ ({\rm Heavy \ Higgs \ scalar \ mass}) = 3000~{\rm GeV},
 \nonumber\\
 &({\rm sfermion \ mass}) = 3000~{\rm GeV}, \ \ \ ({\rm gaugino \ mass}) = 3000~{\rm GeV}.
\end{align}
For the above mass spectrum, the loop functions ${\cal F},\,{\cal F}_\tau,\,{\cal F}'$ are computed as
\begin{align}
{\cal F}={\cal F}_\tau=-1, \ \ \ \ \ {\cal F}'=-0.5
\end{align}
We fix the GUT scale as $\mu_{\rm GUT}=2\times10^{16}$~GeV,
 the soft SUSY breaking scale as $\mu_{\rm SUSY}=3000$~GeV, and the hadronic scale as $\mu_{\rm had}=2$~GeV.
The values of the hadronic form factors are adopted from Ref.~\cite{Aoki:2017puj} as 
 $\alpha_H(2~{\rm GeV})=-\beta_H(2~{\rm GeV})=-0.0144$~GeV.
The values of the baryon chiral Lagrangian parameters and the pion decay constant are
 $D=0.80$, $F=0.46$, $f_\pi=0.093$~GeV.

Yukawa coupling textures at the GUT scale depend on $\tan\beta$ \cite{GGRossFermionMass,GotoRRRR}.
We investigate the cases of $\tan \beta = 3, 10, 30,$ and $50$.
The weak gauge coupling at the scale $\mu_{\rm SUSY}$ is calculated to be $g_2(\mu_{\rm SUSY})=0.634$ for all chosen values of $\tan \beta$.
Table~\ref{tab:yukawa_couplings} summarizes the Yukawa couplings at the GUT scale $\mu_{\rm GUT}$,
 and the top and tau Yukawa couplings at $\mu_{\rm SUSY}$.  
Similarly, Table~\ref{tab:ckm_matrix} shows the CKM matrix elements at $\mu_{\rm GUT}$.
The specific components $V_{ts}$ and $V_{td}$ at $\mu_{\rm SUSY}$ are evaluated as follows:
\begin{align}
    V_{ts}(\mu_{\rm SUSY})=0.0428-0.0008\, i, \quad
    V_{td}(\mu_{\rm SUSY})=0.00826 + 0.00342\, i
\end{align}
\begin{table}[t]
    \centering
    \caption{Yukawa couplings for $\tan\beta = 3, 10, 30, 50$}
    \renewcommand{\arraystretch}{1.2}
    \begin{tabular}{c|cccc}
        \hline\hline
        Coupling & $\tan\beta = 3$ & $\tan\beta = 10$ & $\tan\beta = 30$ & $\tan\beta = 50$ \\
        \hline
        $y_u(\mu_{\rm GUT})$ 
          & $3.00 \times 10^{-6}$ & $2.75 \times 10^{-6}$ & $2.74 \times 10^{-6}$ & $2.77 \times 10^{-6}$ \\
        $y_c(\mu_{\rm GUT})$ 
          & $1.54 \times 10^{-3}$ & $1.41 \times 10^{-3}$ & $1.41 \times 10^{-3}$ & $1.43 \times 10^{-3}$ \\
        $y_t(\mu_{\rm GUT})$ 
          & $0.547$  & $0.483$  & $0.488$   & $0.516$ \\
        \hline
    
        $y_d(\mu_{\rm GUT})$ 
          & $1.55 \times 10^{-5}$ & $4.95 \times 10^{-5}$ & $1.56 \times 10^{-4}$ & $3.00 \times 10^{-4}$ \\
        $y_s(\mu_{\rm GUT})$ 
          & $3.08 \times 10^{-4}$ & $9.86 \times 10^{-4}$ & $3.10 \times 10^{-3}$ & $5.98 \times 10^{-3}$ \\
        $y_b(\mu_{\rm GUT})$
          & $0.0175$     & $0.0555$  & $0.181$  & $0.389$  \\
        \hline
        $y_e(\mu_{\rm GUT})$ 
          & $6.37 \times 10^{-6}$ & $2.03 \times 10^{-5}$ & $6.40 \times 10^{-5}$ & $1.23 \times 10^{-4}$ \\
        $y_\mu(\mu_{\rm GUT})$
          & $1.34 \times 10^{-3}$ & $4.29 \times 10^{-3}$ & $1.35 \times 10^{-2}$ & $2.60 \times 10^{-2}$ \\
        $y_\tau(\mu_{\rm GUT})$ 
          & $0.0229$  & $0.0733$  & $0.240$   & $0.516$   \\
        \hline
        $y_t(\mu_{\rm SUSY})$ & $0.844$ & $0.805$ & $0.801$ & $0.801$ \\
        $y_\tau (\mu _{\rm SUSY})$ & $0.0325$ & $0.103$ & $0.308$ & $0.514$\\
        \hline \hline
    \end{tabular}
    \label{tab:yukawa_couplings}
\end{table}
\begin{table}[t]
    \centering
    \caption{
    CKM matrix elements ($V_{CKM}$) at $\mu_{\rm GUT}$ for $\tan\beta = 3, 10, 30, 50$
    }
    \renewcommand{\arraystretch}{1.2}
    \resizebox{\textwidth}{!}{
    \begin{tabular}{c|cccc}
        \hline\hline
 
        Element & $\tan\beta = 3$ & $\tan\beta = 10$ & $\tan\beta = 30$ & $\tan\beta = 50$ \\
        \hline
        $V_{ud}$ 
          & $-0.400 - 0.889\,i$ & $-0.400 - 0.889\,i$ & $0.0294 + 0.974\,i$  & $0.0295 + 0.974\,i$  \\
        $V_{us}$ 
          & $-0.0922 - 0.205\,i$ & $-0.0922 - 0.205\,i$ & $0.0960 + 0.203\,i$  & $0.0960 + 0.203\,i$  \\
        $V_{ub}$
          & $-0.00353 - 2.47 \times 10^{-6}\,i$ & $-0.00357 - 2.56 \times 10^{-6}\,i$ & $-0.00352 - 2.46 \times 10^{-6}\,i$ & $-0.00339 - 2.18 \times 10^{-6}\,i$ \\
        \hline
        $V_{cd}$
          & $0.225 + 1.40 \times 10^{-4}\,i$ & $0.225 + 1.40 \times 10^{-4}\,i$ & $-0.207 - 0.0860\,i$ & $-0.207 - 0.0860\,i$ \\
        $V_{cs}$ 
          & $-0.974$ & $-0.974$ & $0.973 - 0.0180\,i$ & $0.974 - 0.0179\,i$ \\
        $V_{cb}$ 
          & $-0.0399$ & $-0.0404$ & $-0.0399$ & $-0.0384$ \\
      
        \hline
        $V_{td}$ 
          & $0.00756 - 0.00313\,i$ & $0.00765 - 0.00317\,i$ & $-0.00817$ & $-0.00786$ \\
        $V_{ts}$ 
          & $-0.0392 - 7.23 \times 10^{-4}\,i$ & $-0.0397 - 7.31 \times 10^{-4}\,i$ & $0.0392$ & $0.0377$ \\
        $V_{tb}$ 
          & $0.999$ & $0.999$ & $0.999$ & $0.999$ \\
        \hline \hline
    \end{tabular}
    }
    \label{tab:ckm_matrix}
\end{table}
The RG evolution factors for the Wilson coefficients depend on $\tan \beta$.
Their values evaluated at the relevant energy scales are summarized in Table~\ref{tab:rg_factors}.
\begin{table}[t]
    \centering
    \caption{
    RG evolution factors evaluated for $\tan\beta = 3, 10, 30, 50$
    }
    \renewcommand{\arraystretch}{1.2}
    \begin{tabular}{c|cccc}
        \hline\hline
        RG factor & $\tan\beta = 3$ & $\tan\beta = 10$ & $\tan\beta = 30$ & $\tan\beta = 50$ \\
        \hline
        $A_L^\tau(\mu_{\rm SUSY},\mu_{\rm GUT})$     
          & $4.65$ & $4.64$ & $4.52$ & $4.19$ \\
        $A_R^{\tau t}(\mu_{\rm SUSY},\mu_{\rm GUT})$ 
          & $2.67$ & $2.74$ & $2.66$ & $2.45$ \\
        $A_L(\mu_{\rm SUSY}, \mu_{\rm GUT})$
          & $4.65$ & $4.65$ & $4.65$ & $4.65$ \\
        $A_{LL}(\mu_{\rm had}, \mu_{\rm SUSY})$ 
          & $1.55$ & $1.55$ & $1.55$ & $1.55$ \\
        $A_{LR}(\mu_{\rm had}, \mu_{\rm SUSY})$ 
          & $1.59$ & $1.59$ & $1.59$ & $1.59$ \\
        \hline\hline
    \end{tabular}
    \label{tab:rg_factors}
\end{table}

The free parameters are $U_u,\,U_e,\,V_e$ and $M,\lambda,\kappa,\kappa',\lambda^a,\lambda^b$.
Then $m$ is determined by the condition that ${\rm det}{\cal M}_{\rm doublet}=0$
 (which is equivalent to the condition that ${\cal M}_{\rm doublet}$ have one almost zero singular value).
Their ranges in the numerical analysis are
\begin{align}
&U_u,\,U_e,\,V_e \ : \ {\rm all \ possible \ unitary \ matrices}
\nn\\
&\lambda,\kappa,\kappa',\lambda^a,\lambda^b \ : \ [0.1, \ 1]
\nn\\
&M\ : \ [0.5, \ 5]\times10^{16}~{\rm GeV}.
\label{parameterrange}
\end{align}

The unitary matrices $U_u, U_e,$ and $V_e$ are parameterized by nine real angles $(\phi_0, \phi_1, \phi_2, \theta_1, \theta_2, \delta, \theta_3, \chi_1, \chi_2)$ as in Eq.~(\ref{eq:UnitaryMatrix}). 
In our numerical analysis,
 each of these $3 \times 9 = 27$ angles is randomly sampled from a uniform distribution $x_i \in [0, 2\pi)$.
\begin{equation}
    U(\phi_0,\phi_1,\phi_2,\theta_1,\theta_2,\delta,\theta_3,\chi_1,\chi_2)
    = e^{i\phi_0}\, e^{i(\phi_1\lambda_3+\phi_2\lambda_8)}\,
    R(\theta_1,\theta_2,\delta,\theta_3)\,
    e^{i(\chi_1\lambda_3+\chi_2\lambda_8)},
\label{eq:UnitaryMatrix}
\end{equation}
where $R(\theta_1,\theta_2,\delta,\theta_3)$ is a CKM-like mixing matrix given by
\begin{equation}
    R =
    \begin{pmatrix}
        1 & 0 & 0 \\
        0 & c_1 & s_1 \\
        0 & -s_1 & c_1
    \end{pmatrix}
    \begin{pmatrix}
        c_2 & 0 & s_2 \,e^{-i\delta} 
\\
        0 & 1 & 0 \\
        -s_2 \,e^{i\delta} & 0 & c_2
    \end{pmatrix}
    \begin{pmatrix}
        c_3 & s_3 & 0 \\
        -s_3 & c_3 & 0 \\
        0 & 0 & 1
    \end{pmatrix}.
\end{equation}
Here,
 we adopt the shorthand notations $c_i \equiv \cos\theta_i$ and $s_i \equiv \sin\theta_i$.
The diagonal generators are defined as
\begin{equation}
    \lambda_3 = \mathrm{diag}(1,-1,0), \quad \lambda_8 = \frac{1}{\sqrt{3}}\mathrm{diag}(1,1,-2).
\end{equation}

The value of the GUT-breaking VEV $v$ is assumed to be $v=2.8\times10^{16}$~GeV
 (which corresponds to GUT gauge boson mass of $M_X\simeq2\times 10^{16}$~GeV and the SU(5) gauge coupling at scale $\mu_{\rm GUT}$ of 0.7).
We optimize the model parameters to maximize the proton lifetime.
This is achieved by minimizing the following loss function,
 which represents the mean log-scaled partial width of $p \to K^+ \bar \nu$
 (see Eqs. (\ref{nutau}), (\ref{numue}) ):
\begin{align}
    \text{Loss function} 
    = \frac{1}{N} \sum_{i=1}^{N} 
    \log_{10} \Gamma(p \to K^+ \bar{\nu})_i,
    \label{eq:LossFunction}
\end{align}
where $N$ denotes the number of parameter configurations and $\Gamma(p \to K^+ \bar{\nu})$ is expressed in GeV.
This loss function is minimized over $N_{\rm iter}$ iterations using an optimization algorithm with a batch size of $N_{\rm batch}$.
In each iteration, 
 the $N$ parameter configurations are randomly shuffled and divided into $N/N_{\rm batch}$ mini-batches.
The loss function and its gradient are evaluated on each mini-batch in turn,
 and the parameters are updated using the gradient of that mini-batch,
 so that every configuration contributes to the update once per iteration.
As a result, 
 we obtain a minimized loss function and the corresponding set of optimized parameters, $x_i$.

\section{Numerical Results}
\label{section-Numerical-Results}

In this analysis,
 large-scale numerical computations and the optimization procedure were implemented using the machine learning library TensorFlow~\cite{TensorFlow}.
We randomly generated $N=4096$ initial configurations of 33 parameters from a uniform distribution.
The parameters were then optimized to minimize the loss function Eq.(\ref{eq:LossFunction}) for up to 100,000 iterations.
We employed the Adam optimizer \cite{Adam} with a learning rate of 0.001,
 while the other hyperparameters were set to their default values.
The Adam optimizer iteratively updated the $N=4096$ parameter configurations with a batch size of $N_{\rm batch} = 1024$.

\begin{figure}[H] 
    \centering

    \includegraphics[scale=0.6]{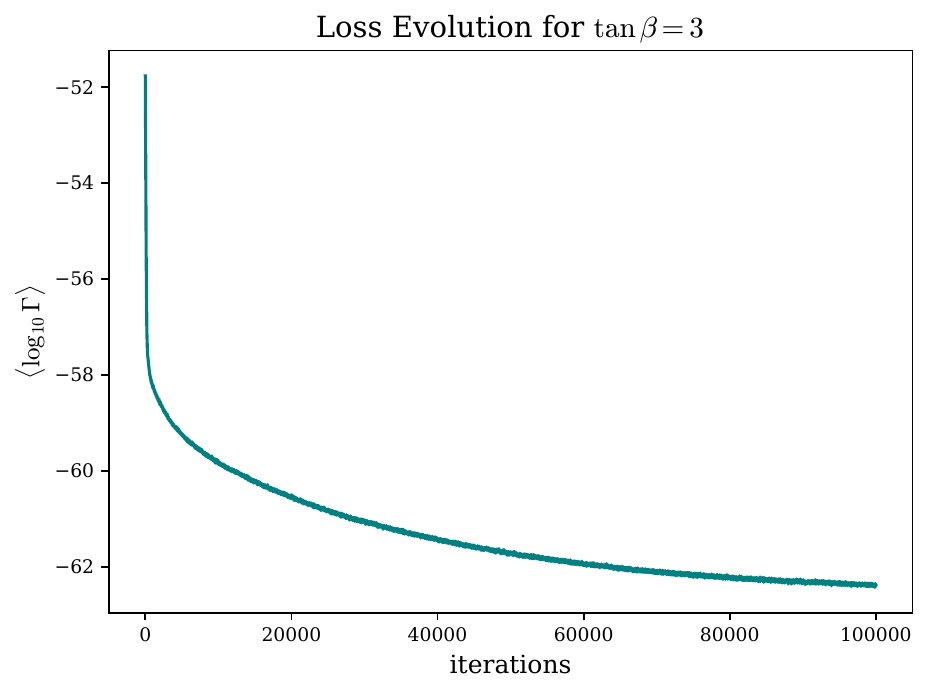}
    
    \vspace{0.5cm}

    \includegraphics[scale=0.6]{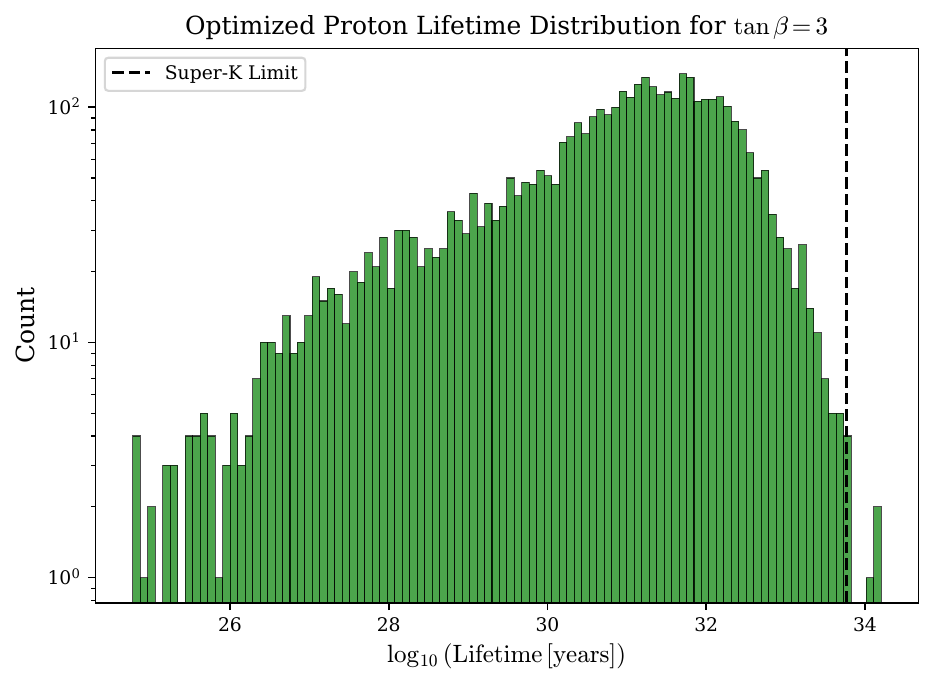}

    \caption{
    Optimization results for the proton decay suppression for $\tan\beta=3$.
    The top panel shows the evolution of the loss function over the iterations.
    The bottom panel displays the distribution of the proton lifetimes for the optimized $N=4096$ samples, categorized into 100 bins.
    The vertical dashed line in the bottom panel indicates the experimental lower bound of $5.9 \times 10^{33}$ years from Super-K
\cite{Super-K-Limit}.
}
    \label{fig:optimization_tanb3}
\end{figure}

Figure \ref{fig:optimization_tanb3} summarizes the numerical optimization process for $\tan \beta =3$.
The top panel shows that the loss function rapidly decreases,
 indicating that the Adam optimizer efficiently explores the high-dimensional parameter space.
The bottom panel shows the proton lifetime distribution of the $N=4096$ parameter sets after optimization.
While the initial random configurations mostly reside below the experimental bound, 
 the optimized samples are successfully pushed toward the Super-K limit of $5.9 \times 10^{33}$ years \cite{Super-K-Limit}.

\begin{figure}[p] 
    \centering
    
    \begin{minipage}{\textwidth}
        \centering
        \includegraphics[width=\textwidth]{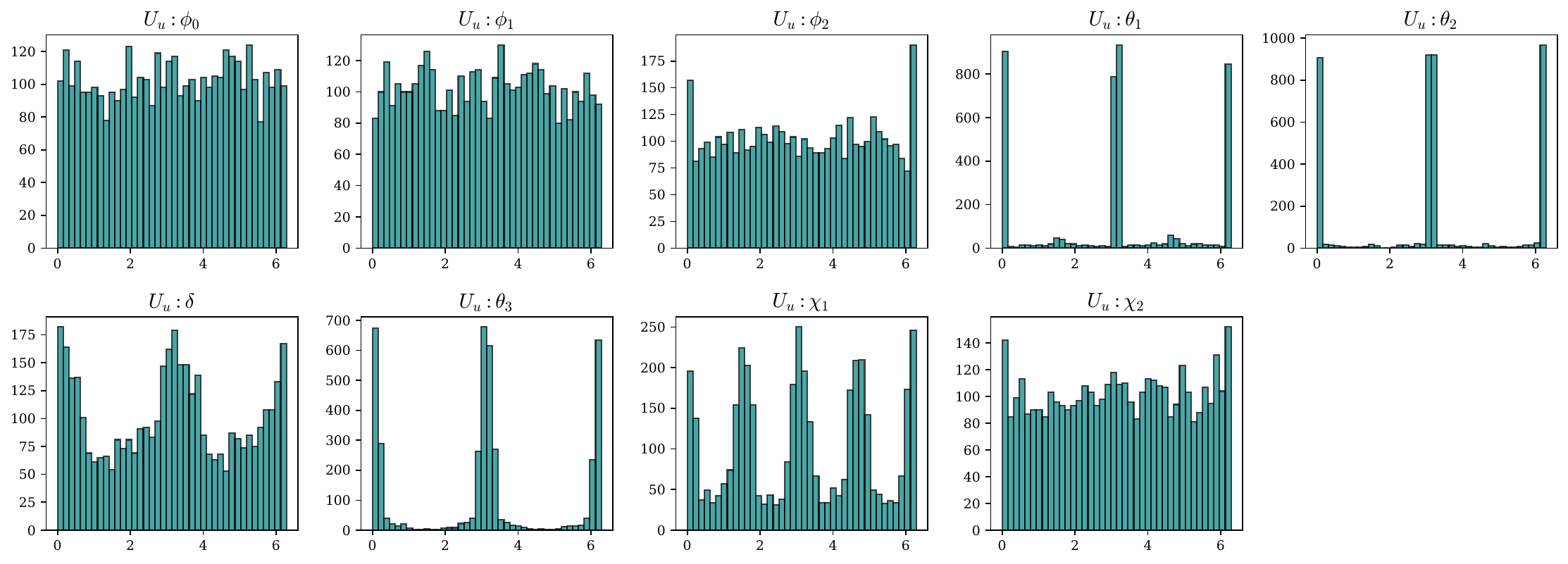}
        \vspace{2pt} \\ (a)
    \end{minipage}
    
    \vspace{2em} 
    
    \begin{minipage}{\textwidth}
        \centering
        \includegraphics[width=\textwidth]{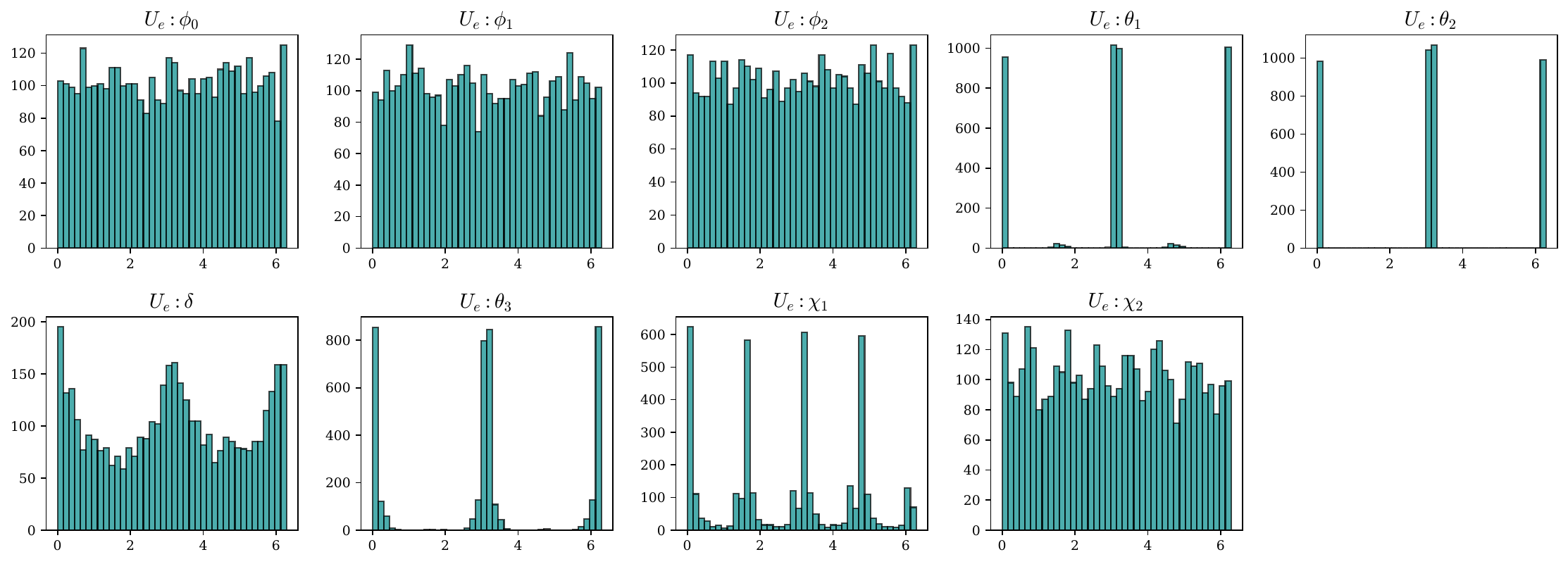}
        \vspace{2pt} \\ (b)
    \end{minipage}
    
    \caption{
        Distributions of the $N=4096$ optimized parameters for $\tan\beta=3$ (40 bins each).
        (a) The 9 mixing angles ($\phi_i, \theta_i, \delta, \chi_i$) for the unitary matrix $U_u$. 
        (b) The 9 mixing angles for the unitary matrix $U_e$. 
    }
    \label{fig:params_dist_tanb3_part1}
\end{figure}

During the optimization, 
 the parameters $x_i$ navigate the parameter space toward regions where the loss function takes smaller values.
The distributions of the $N=4096$ optimized parameters are detailed in Fig. \ref{fig:params_dist_tanb3_part1}.

\addtocounter{figure}{-1}

\begin{figure}[p] 
    \centering
    
    \begin{minipage}{\textwidth}
        \centering
        \includegraphics[width=\textwidth]{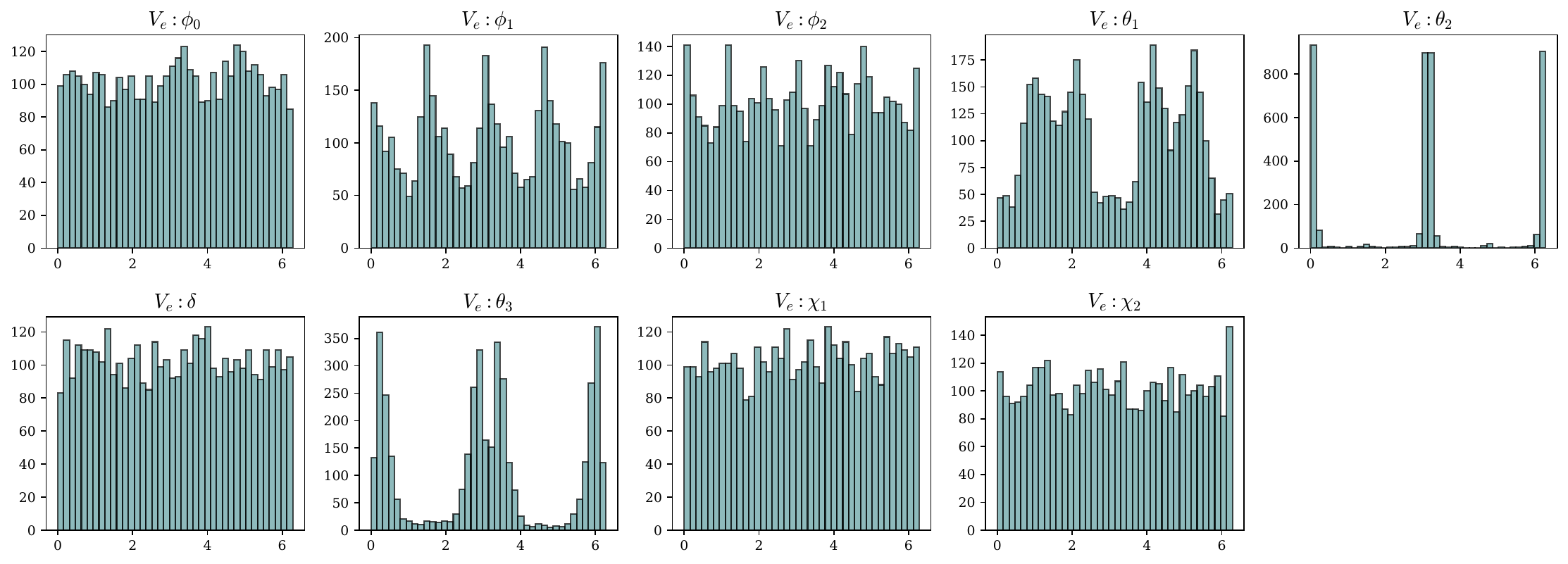}
        \vspace{2pt} \\ (c)
    \end{minipage}
    
    \vspace{2em} 
    
    \begin{minipage}{\textwidth}
        \centering
        \includegraphics[width=\textwidth]{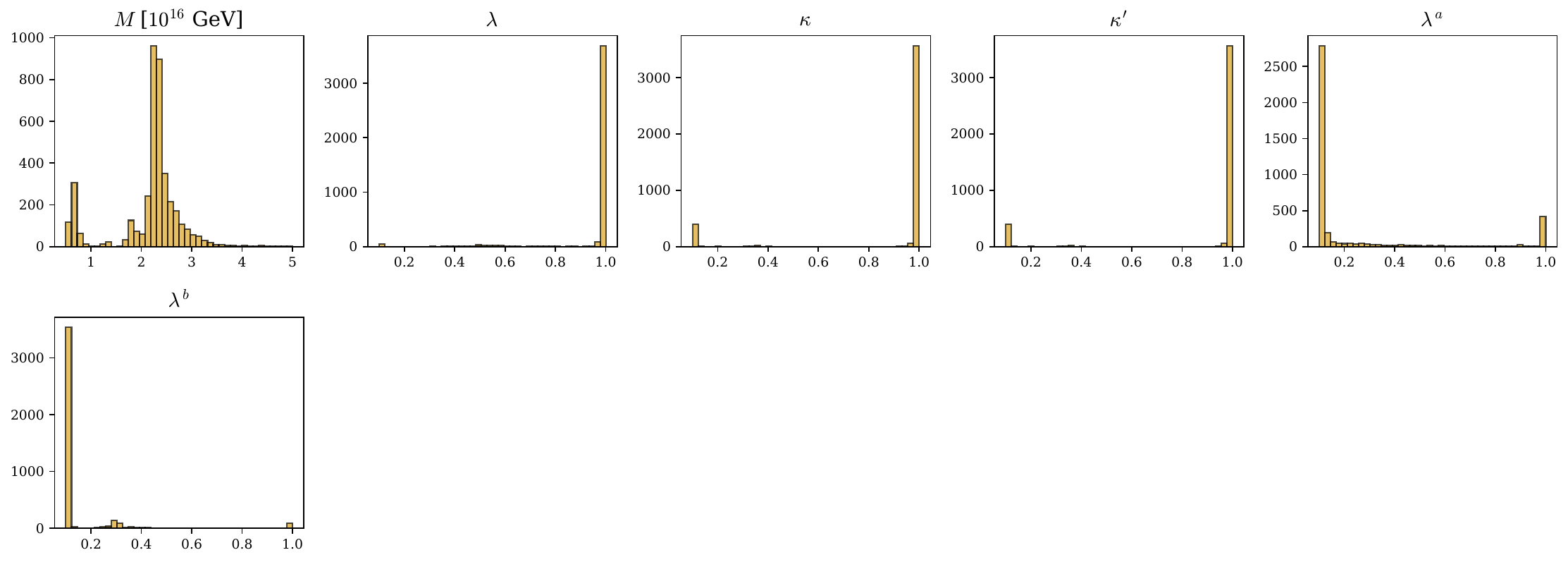}
        \vspace{2pt} \\ (d)
    \end{minipage}
    
    \caption{
        (Continued) 
        (c) The 9 mixing angles for the unitary matrix $V_e$. 
        (d) The 6 potential parameters ($M, \lambda, \kappa, \kappa', \lambda^a, \lambda^b$).
    }
    \label{fig:params_dist_tanb3_part2}
\end{figure}

It is observed that several mixing angles and potential couplings, such as $M$ and $\kappa$, exhibit non-trivial localized distributions rather than remaining uniform.
This localization suggests that the suppression of proton decay requires a specific alignment between the Yukawa matrices and the potential sector parameters.

\begin{figure}
    \centering
    \includegraphics[width=\linewidth]{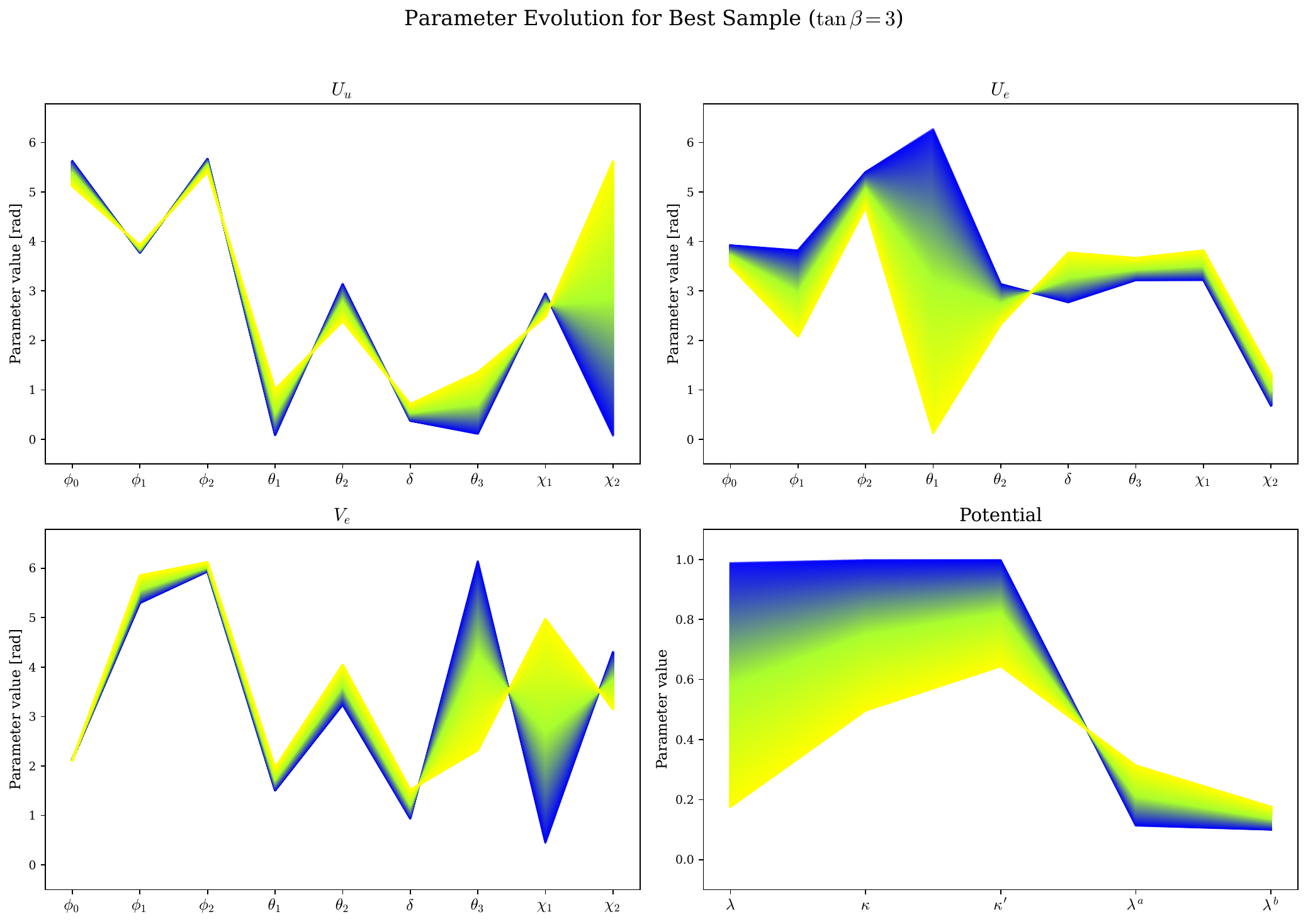}
    \caption{
    Evolution of the optimal parameters that suppress proton decay for $\tan\beta = 3$ 
    (the parameter $M$ is optimized from its initial random value of $3.77 \times 10^{16}$ GeV to $2.25 \times 10^{16}$ GeV).
    The brightest yellow lines and the darkest blue lines in the panels represent the initial random values ($N_{\rm iter}=0$) and the optimized configuration at $N_{\rm iter}=100,000$, respectively.
      }
    \label{fig:BestParaEvol}
\end{figure}

\begin{figure}[t]
    \centering
    \includegraphics[width=\linewidth]{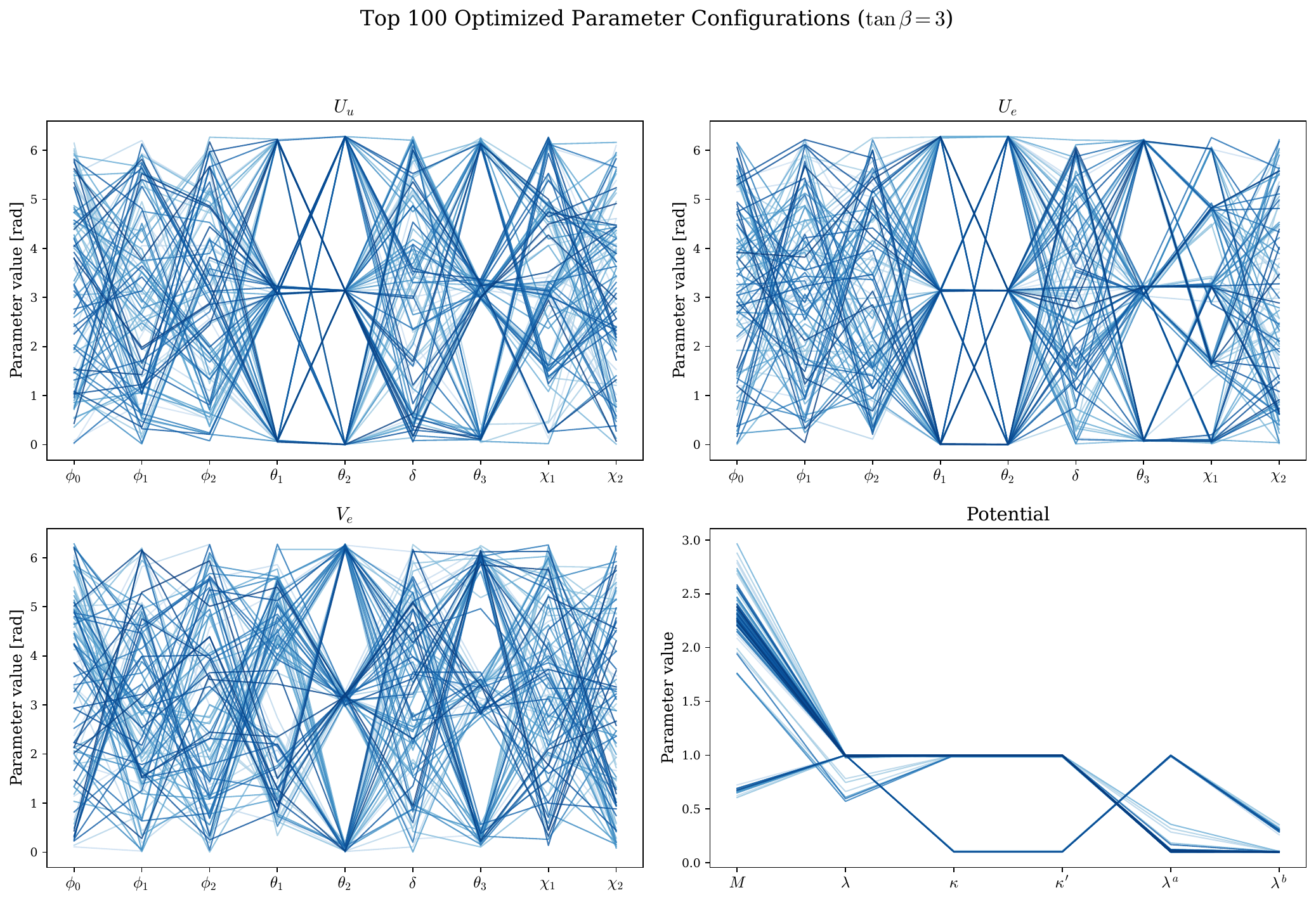}
    \caption{ 
    The top 100 configurations of the optimal parameters that suppress proton decay for $\tan\beta = 3$. 
    Note that $M \sim \mathcal{O}(10^{16}\, {\rm GeV})$ in the analysis, while $M$ is normalized in the figure. 
    Darker lines correspond to smaller values of the optimized loss function.
      } 
    \label{fig:top100}
\end{figure}

Fig.~\ref{fig:BestParaEvol} shows the parameter evolution of the best sample for suppressing proton decay at $\tan\beta=3$.
Some parameters undergo large changes from their initial values, 
 while others remain close to themselves. 
This is consistent with a picture in which the loss landscape contains both steep and shallow directions: 
 the former would account for the rapid initial drop in Fig.~\ref{fig:optimization_tanb3},
 and the latter for the subsequent slow decrease. 
 In particular, 
 the potential parameters tend to favor large $(\sim 1)$ or small $(\sim  0.1)$ values,
 as can be seen in Fig.~\ref{fig:params_dist_tanb3_part2}.
This tendency may suggest the presence of steep directions in the loss landscape associated with the potential parameters.
Fig.~\ref{fig:top100} shows the top 100 optimized parameter configurations for suppressing proton decay at $\tan \beta = 3$.
As illustrated in Fig.~\ref{fig:params_dist_tanb3_part1},
 the optimized angles $\theta_i\, (i=1,2,3)$ of each CKM-like matrix and the optimized potential parameters tend to several specific values.

\begin{figure}[p]
    \centering
    \begin{minipage}{0.48\textwidth}
        \centering
        \includegraphics[width=0.9\textwidth]{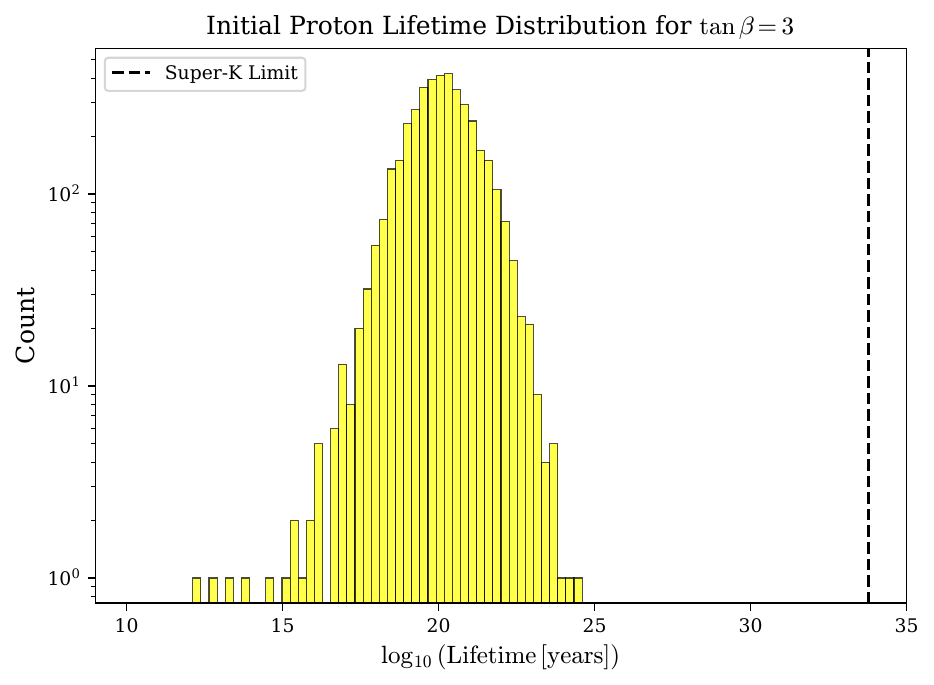}
        \vfill
    \end{minipage}
    \hfill
    \begin{minipage}{0.48\textwidth}
        \centering
        \includegraphics[width=0.9\textwidth]{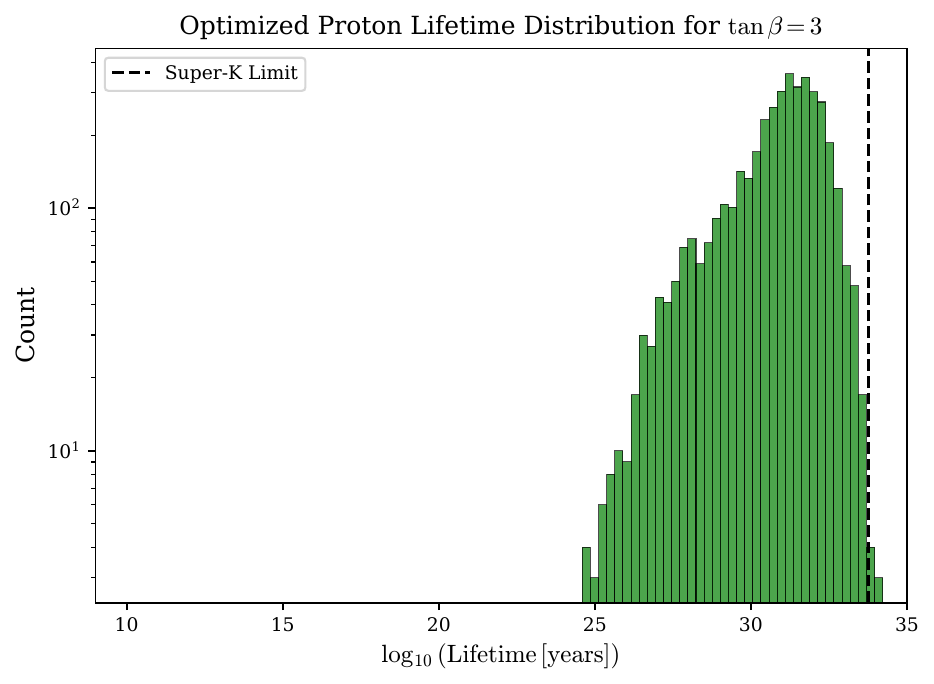} 
        \vfill
    \end{minipage}


    \begin{minipage}{0.48\textwidth}
        \centering
        \includegraphics[width=0.9\textwidth]{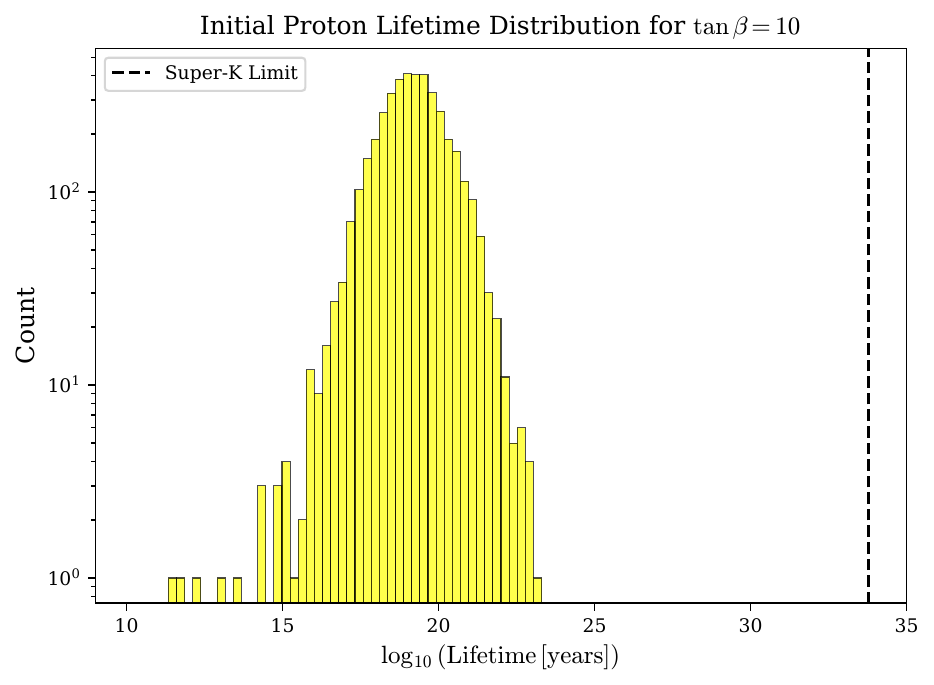}
        \vfill
    \end{minipage}
    \hfill
    \begin{minipage}{0.48\textwidth}
        \centering
        \includegraphics[width=0.9\textwidth]{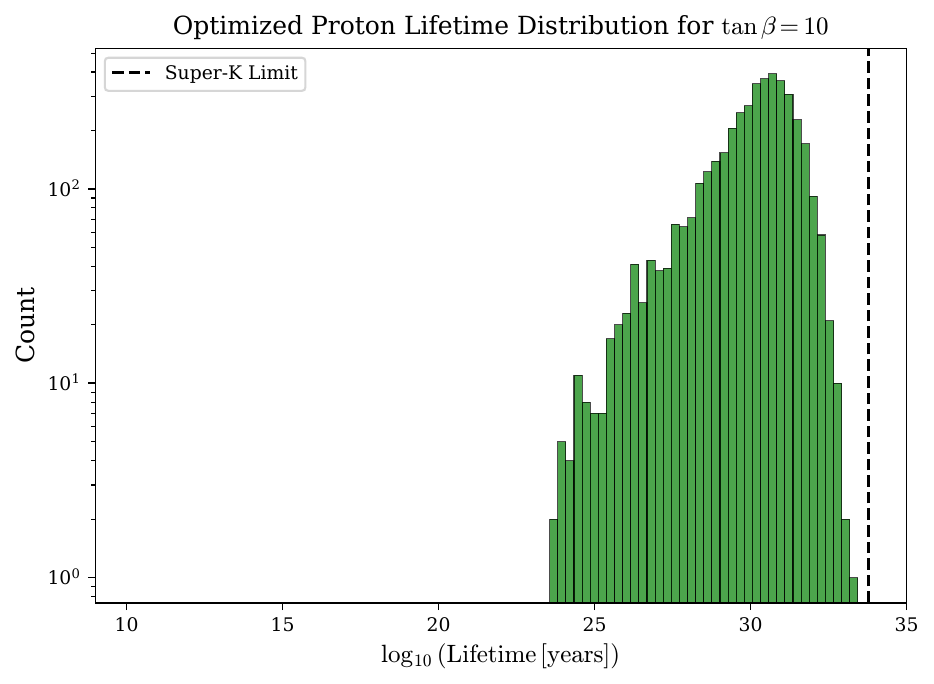}
        \vfill
    \end{minipage}


    \begin{minipage}{0.48\textwidth}
        \centering
        \includegraphics[width=0.9\textwidth]{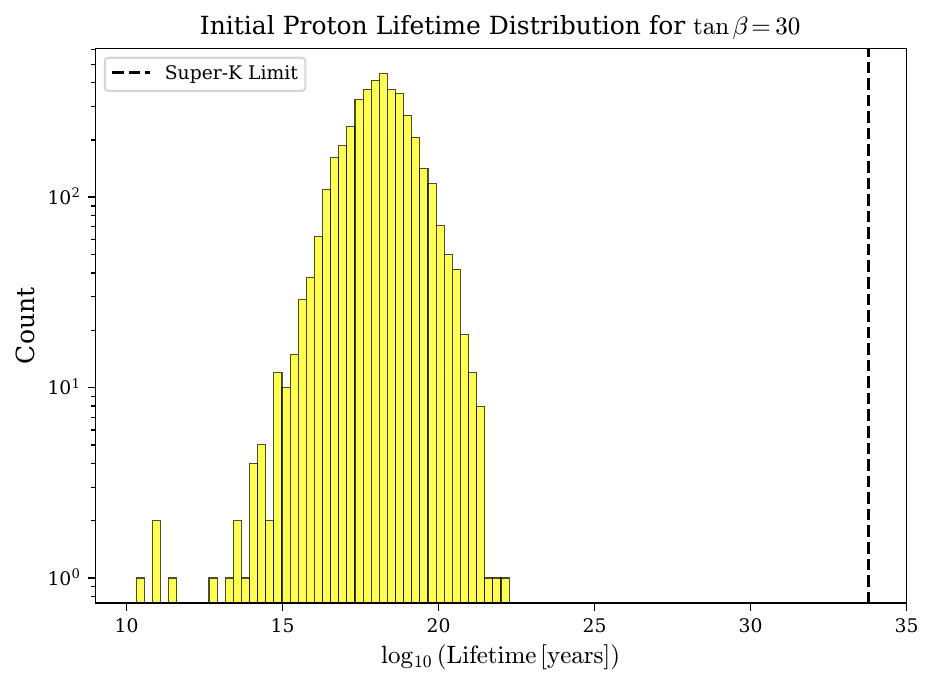}
        \vfill
    \end{minipage}
    \hfill
    \begin{minipage}{0.48\textwidth}
        \centering
        \includegraphics[width=0.9\textwidth]{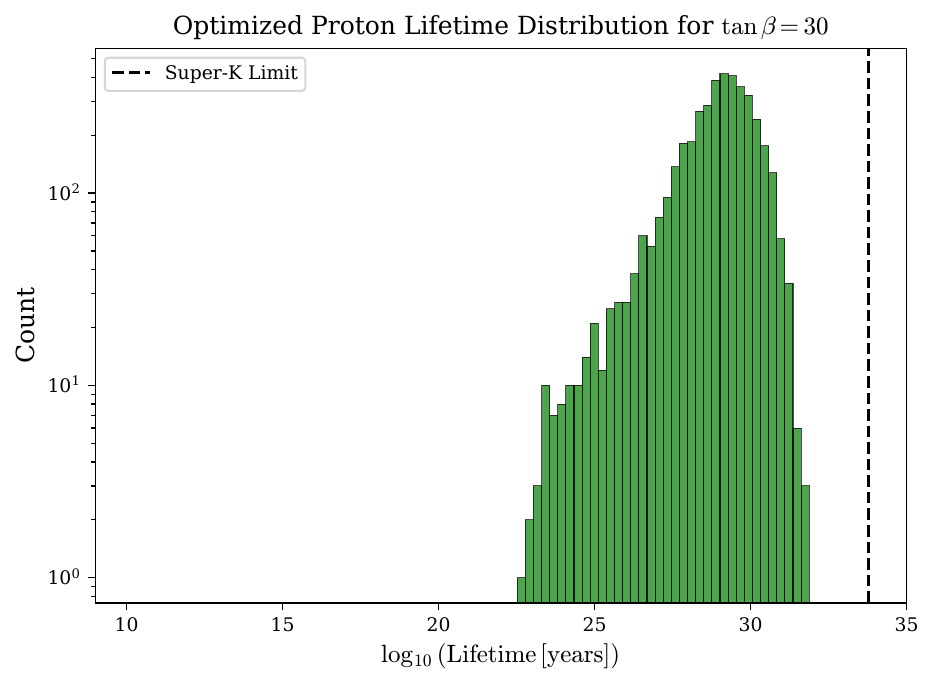}
        \vfill
    \end{minipage}


    \begin{minipage}{0.48\textwidth}
        \centering
        \includegraphics[width=0.9\textwidth]{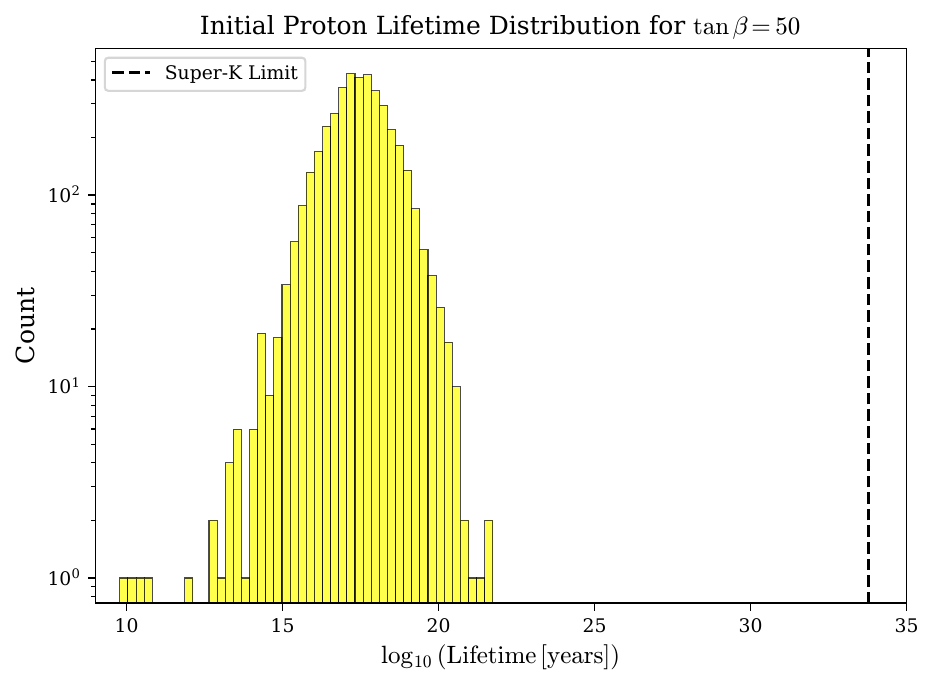}
        \vfill
    \end{minipage}
    \hfill
    \begin{minipage}{0.48\textwidth}
        \centering
        \includegraphics[width=0.9\textwidth]{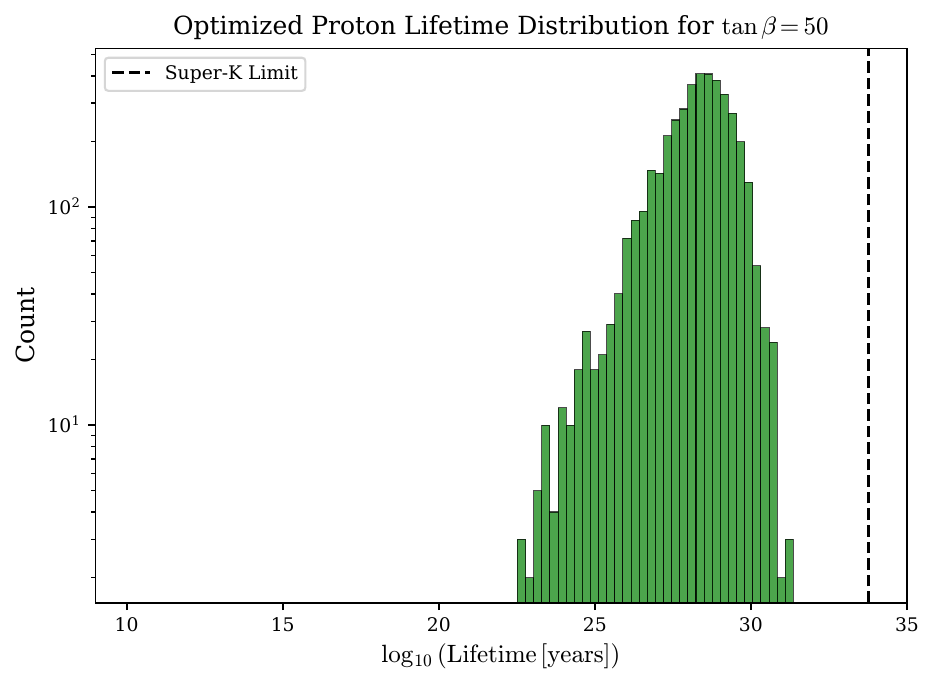}
        \vfill
    \end{minipage}

    \caption{
Comparison of the initial and optimized proton lifetime distributions for $\tan\beta = 3, 10, 30,$ and $50$. 
The left (right) panels display the distributions for the initial (optimized) $N=4096$ samples with 100 bins.
}
    \label{fig:initial VS Optimzed lifetime}
\end{figure}

Finally, the proton lifetime distributions for the initial random configurations and the optimized samples for $\tan\beta = 3, 10, 30,$ and $50$ are summarized in Fig. \ref{fig:initial VS Optimzed lifetime}.
A comparison between these two figures illustrates how the optimization of the parameters $x_i$ consistently shifts the distributions toward longer lifetimes and suppresses the $p \to K^+ \bar{\nu}$ decay rate. 
Note that the horizontal-axis scales in Fig.~\ref{fig:initial VS Optimzed lifetime} differ from that in Fig.~\ref{fig:optimization_tanb3}.%
While a significant portion of the configurations remains below the experimental bound,
 the optimizer successfully identifies specific parameter regions where the proton lifetime is substantially enhanced.

\section{Summary}
\label{section-summary}

In this paper, 
 we addressed the formidable challenge of rapid dimension-five proton decay in the supersymmetric $SU(5)$ GUT.
Recognizing that the enlarged Yukawa and potential sectors—required to correctly reproduce the SM fermion mass spectrum—introduce a high-dimensional parameter space,
 we leveraged machine learning techniques to systematically investigate regions consistent with current experimental bounds.

By defining an objective loss function based on the theoretically predicted $p \to K^+ \bar{\nu}$ partial decay width,
 we utilized the Adam optimization algorithm to explore a 33-dimensional space consisting of complex mixing angles, phases, and potential sector parameters.
We evaluated $N = 4096$ randomly generated initial configurations across several values of $\tan \beta$ (3, 10, 30, and 50). 
Our study demonstrates that the optimizer can efficiently navigate away from random parameter regions that severely violate proton decay bounds. 
While not all configurations reach the experimentally allowed regime,
 the algorithm successfully identifies non-trivial parameter regions where the proton lifetime is substantially enhanced,
 pushing a fraction of the samples beyond the current Super-K limit of $5.9 \times 10^{33}$ years.

Crucially, our analysis reveals that the optimized parameters do not populate the allowed parameter space uniformly;
 rather, they exhibit a distinct localization behavior.
These results indicate that achieving sufficient suppression of dimension-five proton decay necessitates highly specific, 
 non-trivial alignments between the Yukawa matrices and the GUT-scale potential sector parameters. 
In our model,
 the suppression of proton decay is primarily driven by the flavor mixing structure,
 characterized by a CKM-like matrix $R(\theta_i, \delta)$. 
The optimization process automatically identifies specific mixing angles and phases that minimize the triplet-Higgsino-mediated operators.
However, 
 we find that the achievable proton lifetime is highly sensitive to the value of $\tan \beta$. 
Our results for $\tan \beta = 3, 10, 30, \text{and } 50$ demonstrate a clear trend: 
 the optimized proton lifetime decreases significantly as $\tan \beta$ increases. 

The optimization results reveal a strong correlation between $\tan \beta$ and proton stability.
While our machine learning approach successfully identifies localized regions with suppressed decay widths,
 the optimized proton lifetime exhibits a sharp sensitivity to the value of $\tan \beta$.
Specifically, 
 as $\tan \beta$ increases, 
 the enhancement of down-type quark and lepton sector contributions significantly amplifies the dimension-five proton decay operators.
This intrinsic enhancement creates a formidable challenge for achieving sufficient suppression,
 as it necessitates increasingly precise and non-trivial cancellations within the 33-dimensional parameter space.
Our analysis suggests that in the high $\tan \beta$ regime, 
 the required suppression for triplet-Higgsino-mediated decay increasingly strains the model's capacity to meet the stringent Super-K lower bounds.
We note that the $3$~TeV benchmark spectrum adopted here lies above the current LHC limits on gluinos and squarks~\cite{ATLAS:2020syg,CMS:2019zmd},
 and that $\tan\beta = 3$ is not excluded by the heavy Higgs searches in the $\tau\tau$ channel,
 whose sensitivity is limited to large $\tan\beta$ and decreases with the heavy Higgs mass~\cite{ATLAS:2020zms,CMS:2022goy}.

This study demonstrates the effectiveness of machine learning approaches in addressing challenging phenomenological problems in beyond-the-SM physics,
 particularly those associated with the curse of dimensionality.
The framework developed in this work provides a flexible and powerful approach that can be extended to other GUTs, 
 such as $SO(10)$, as well as to scenarios involving additional stringent phenomenological constraints.
We have focused on the $p \to K^+ \bar\nu$ mode throughout this work.
The next-leading modes $p \to \pi^+ \bar\nu$ are suppressed by the Cabibbo angle relative to the $K\bar\nu$ modes,
 and the modes with charged leptons are further suppressed by the small up-quark Yukawa coupling in the minimal model,
 as shown in the analysis based on the chiral Lagrangian technique in Ref.~\cite{Hisano:1992jj}.
Nevertheless, a complete analysis should also include the $\pi^+ \bar\nu$ modes,
 and a systematic study of these modes is left for future work.

\section*{Acknowledgement}
This work is partially supported by Scientific Grants by the Ministry of Education, Culture, Sports, Science and Technology of Japan
 No.~23K03392 (N.H.).


\end{document}